\documentclass[aps,nofootinbib,superscriptaddress,10pt,floatfix]{revtex4}
\normalsize
\usepackage{mathrsfs}
\usepackage{graphicx}
\usepackage{amsmath,amssymb}
\usepackage{braket}
\usepackage{subfigure}
\usepackage{subfloat}
\usepackage{float}
\usepackage[usenames]{color}
\usepackage[normalem]{ulem}
\usepackage{comment}
\usepackage{ifthen} 
\usepackage{fp,calc}
\usepackage{lastpage}
\usepackage{setspace}
\usepackage[pdftex]{epsfig}
\DeclareGraphicsExtensions{.jpg,.mps,.pdf,.png}
\usepackage{cellspace}
\usepackage{booktabs}
\usepackage{latexsym}
\usepackage{graphics}
\usepackage{fancyhdr}
\usepackage{adjustbox}
\usepackage[english]{babel}
\usepackage[svgnames]{xcolor}
\usepackage{enumitem}
\usepackage{tabularx}
\usepackage{multirow}
\usepackage[flushleft]{threeparttablex}
\usepackage{ragged2e}
\usepackage[colorlinks=true, allcolors=blue]{hyperref}

\newcolumntype{C}[1]{>{\centering\arraybackslash}p{#1}}

\usepackage{titlesec}

\newfloat{supfigure}{thp}{losf}
\floatname{supfigure}{Supplementary Figure}
\restylefloat{supfigure}

\newfloat{suptable}{thp}{lost}
\floatname{suptable}{Supplementary Table}
\newfloat{supeq}{thp}{lose}
\floatname{supeq}{Supplementary Equation}
\titleformat{\section}
  {\normalfont\fontsize{12}{14}\bfseries}{\thesection}{1em}{}
\titleformat{\subsection}
  {\normalfont\fontsize{10}{12}\bfseries}{\thesubsection}{1em}{}

\titlespacing*{\section}{0pt}{*2}{*1}
\titlespacing*{\subsection}{0pt}{*2}{*1}

\usepackage{ifthen} 
\usepackage{fp,calc}
\usepackage{lastpage}
\IfPackageLoadedTF{color}{}{\usepackage{color}}

\usepackage{fancyhdr}
\usepackage{adjustbox}
\usepackage[english]{babel}

\IfPackageLoadedTF{xcolor}{}{\usepackage[svgnames, table]{xcolor}}
\usepackage{layout}
\usepackage{enumitem}
\usepackage[normalem]{ulem}
\usepackage{parskip}
\usepackage{multirow}
\usepackage{comment}
\usepackage{booktabs}
\usepackage[flushleft]{threeparttablex}

\newboolean{articletitles}
\setboolean{articletitles}{true} 
\newboolean{allauthors}
\setboolean{allauthors}{false} 

\newcommand{\hJ}{\hat{J}}
\newcommand{\hf}{\hat{f}}
\newcommand{\HH}{\mathcal{H}}
\newcommand{\bV}{\mathbf{V}}
\newcommand{\bk}{\mathbf{k}}

\newcommand{\suppsectionname}{Supplementary Discussion}
\newcommand{\suppeqname}{Supplementary Equation}

\newcommand{\suppref}[2]{%
  \hyperref[#1]{#2}%
} 

\makeatletter
\renewcommand\appendix{%
  \par
  \setcounter{section}{0}%
  \setcounter{subsection}{0}%
  \setcounter{equation}{0}%
  \gdef\thesection{\suppsectionname~\arabic{section}}%
  \gdef\theHsection{supp.\arabic{section}}%
  \gdef\thesubsection{\thesection.\arabic{subsection}}%
  \gdef\theHsubsection{supp.\arabic{section}.\arabic{subsection}}%
  \gdef\theequation{\suppeqname\ \arabic{equation}}%
  \gdef\theHequation{supp.eq.\arabic{equation}}%
}
\makeatother

\begin{document}

\title{Comparing the motion of dark matter and standard model particles\\ on cosmological scales}

\author{Nastassia Grimm}
\email{nastassia.grimm@port.ac.uk}
\affiliation{%
D\'epartement de Physique Th\'eorique and Center for Astroparticle Physics, Universit\'e de Gen\`eve, Quai E. Ansermet 24, CH-1211 Geneva 4, Switzerland}%
\affiliation{%
Institute of Cosmology \& Gravitation, University of Portsmouth, Portsmouth, PO1 3FX, United Kingdom}%
\author{Camille Bonvin}
\email{camille.bonvin@unige.ch}
\affiliation{%
D\'epartement de Physique Th\'eorique and Center for Astroparticle Physics, Universit\'e de Gen\`eve, Quai E. Ansermet 24, CH-1211 Geneva 4, Switzerland}%
\author{Isaac Tutusaus}
\email{tutusaus@ice.csic.es}
\affiliation{Institute of Space Sciences (ICE, CSIC), Campus UAB, Carrer de Can Magrans, s/n, 08193 Barcelona, Spain}
\affiliation{Institut d'Estudis Espacials de Catalunya (IEEC), Edifici RDIT, Campus UPC, 08860 Castelldefels (Barcelona), Spain}
\affiliation{
 Institut de Recherche en Astrophysique et Plan\'etologie (IRAP), Universit\'e de Toulouse, CNRS, UPS, CNES, 14 Av.~Edouard Belin, 31400 Toulouse, France}%

\date{\today}

\begin{abstract}
Since dark matter particles have never been directly detected, we do not know how they move, and in particular we do not know how they fall inside gravitational potential wells. Usually it is assumed that dark matter only interacts gravitationally with itself and with particles of the standard model, and therefore that its motion is governed by Euler's equation. In this paper, we do test this assumption directly at cosmological scales, by combining measurements of galaxy velocities with measurements of gravitational potential wells, encoded in the Weyl potential. We find that current data are consistent with Euler's equation at redshifts $z\in [0.3,0.8]$, and we place constraints on the strength of a potential fifth force, which would alter the way dark matter particles fall. 
We find that a positive fifth force cannot exceed 7\% of the gravitational interaction strength, while a negative fifth force is limited to 21\%. The coming generation of surveys, including the Legacy Survey of Space and Time of the Vera C. Rubin Observatory and the Dark Energy Spectroscopic Instrument will drastically improve the constraints, allowing to constrain a departure from pure gravitational interaction at the level of 2\%.   
\end{abstract}

\maketitle


\section{Introduction}

One of the current main challenges of cosmology and of particle physics is to understand the nature and properties of dark matter. In the simplest model, dark matter is made of cold collisionless particles, which 
interact only gravitationally with particles of the standard model.  This so-called ``cold dark matter'' feels gravity in the same way as standard matter, i.e., it moves along the same geodesics and obeys Euler's equation. The existence of such cold dark matter particles is supported by cosmological observations over a wide range of scales, from the motion of stars in galaxies and that of galaxies in clusters~\cite{Bertone:2016nfn, Salucci:2018hqu}, to the large-scale structure of the Universe~\cite{Blumenthal:1984bp,Davis:1985rj} and the temperature fluctuations of the Cosmic Microwave Background  (CMB)~\cite{Peebles:1982ff,2013ApJS..208...19H,Planck:2018vyg}.

However, since no direct observation of such a particle has been made yet, it is legitimate to question these assumptions and explore models beyond the cold dark matter paradigm. In particular, it is important to test the validity of Euler's equation for dark matter to determine if it falls indeed in the same way inside a gravitational potential well as standard matter. 
A violation of Euler's equation for dark matter can either be gravitational, i.e.,\ due to a breaking of the weak equivalence principle within gravity, which could couple differently to different types of matter, see e.g.,~\cite{Gleyzes:2015pma}. Or it can be due to non-gravitational dark matter interactions, either with particles of the standard model~\cite{Barkana:2018lgd,Schewtschenko:2015rno,Diacoumis:2018ezi}, with a dark sector, e.g.,\ dark radiation~\cite{Diamanti:2012tg,Buen-Abad:2015ova} or dark energy~\cite{Pettorino_2012,Pourtsidou:2013nha,Costa:2014pba}, or with themselves~\cite{Spergel:1999mh,Archidiacono:2022iuu,Tulin:2017ara}. Such interactions would directly alter the way dark matter particles fall in a gravitational potential and break the validity of Euler's equation. In this work we focus on the second scenario: we assume that gravity is described by general relativity, which obeys the weak equivalence principle, and we search for deviations in Euler's equation due to non-gravitational dark matter interactions.

Extensive searches for dark matter particles and their (non-gravitational) interactions have been performed via various methods: searches for dark matter collisions with particles of the standard model in Earth-based detectors, called direct search experiments, e.g.,~\cite{Behnke:2016lsk,CRESST:2019jnq,DarkSide-50:2022qzh,LZ:2022lsv}; searches of products of dark matter decays by looking for new signals from the cosmos, called indirect search experiments, e.g.,~\cite{Gaskins:2016cha,Conrad:2017pms}; and searches of dark matter particles in colliders, in particular at the Large Hadron Collider, see~\cite{Rodriguez:2021urv,CMS:2024zqs,ATLAS:2024lda,ATLAS:2024fdw,FASER:2018eoc}. Since non-gravitational dark matter interactions (if they exist) are believed to be mediated by new particles, colliders are also searching for traces of these new particles, e.g.,~\cite{PhysRevLett.124.041801,ATLAS:2024itc}. In addition, non-gravitational dark matter interactions can be studied through their impact on the formation and evolution of astrophysical objects, such as galaxies and galaxy clusters. Self-interacting dark matter would indeed lead to a non-trivial signature in the central core of dark matter halos~\cite{RobertsonBAHAMAS,Eckert:2022qia}, in the alignment of galaxies~\cite{harveyIA} and in their morphology~\cite{Desmond:2020gzn, Kesden:2006zb}. In this work, we take a complementary approach, and we search for dark matter interactions at cosmological scales by probing the validity of Euler's equation for galaxies. Since galaxies are mainly made of dark matter, by testing the relation between the velocity of galaxies and the gravitational potential $\Psi$, we can directly detect if dark matter particles are subject to a new force.

Cosmological surveys provide measurements of the galaxy peculiar velocities, through the so-called redshift-space distortions~\cite{Kaiser:1987qv,Hamilton:1997zq}. The gravitational potential $\Psi$ has however never been measured at cosmological scales. On the other hand, the Weyl potential, which is the sum of the time distortion $\Psi$ and the spatial distortion $\Phi$, $\Psi_W\equiv (\Phi+\Psi)/2$, 
has recently been measured using gravitational lensing data in a novel way~\cite{Tutusaus:2023aux}. This approach allowed for a direct measurement of $\Psi_W$ at different redshifts. Since in general relativity, the time and spatial distortions are predicted to be the same at late time ($\Phi=\Psi$), we can use measurements of the Weyl potential $\Psi_W=\Psi$ to test the validity of Euler's equation.  

Note that if general relativity is not valid, our test will not hold, since in this case the Weyl potential $\Psi_W$ may differ from the time distortion $\Psi$, and we cannot use it to test the validity of Euler's equation. Methods have been proposed to overcome this limitation~\cite{Bonvin:2018ckp,Castello:2024lhl}, either by measuring directly the distortion of time~\cite{Sobral-Blanco:2022oel}, or through alternative methods, e.g.\ by testing the consistency relations between $(n+1)$-points and $n$-points correlators~\cite{Kehagias:2013rpa,Creminelli:2013nua}. These require however the next generation of data, including the Dark Energy Spectroscopic Instrument (DESI)~\cite{DESI2025}, the Euclid satellite~\cite{Euclid2025} and the Square Kilometer Array Observatory (SKAO)~\cite{SKA2025}. 

In this work, we assume that general relativity is valid, and we test for the presence of non-gravitational interactions (often called fifth force) acting on dark matter in this framework. We constrain the amplitude of the fifth force by combining redshift-space distortions with gravitational lensing. We use recent measurements of the Weyl potential at four different redshifts~\cite{Tutusaus:2023aux}, obtained from galaxy-galaxy lensing and galaxy clustering measurements from the first three years of Dark Energy Survey (DES) data~\cite{DES:2021wwk}, and combine them with measurements of galaxy velocities (encoded in the growth rate of structure) at 22 redshifts from various spectroscopic surveys~\cite{Howlett:2017asq,Huterer:2016uyq,Hudson:2012gt,Turnbull:2011ty,Davis:2010sw,Song:2008qt,Blake:2013nif,eBOSS:2020yzd,Blake:2012pj,Pezzotta:2016gbo,Okumura:2015lvp,eBOSS:2018yfg}. With this we place constraints on the strength of the fifth force at the first four redshift bins considered for the lens galaxies in the DES Year~3 analysis~\cite{DES:2021wwk}. We find that the parameter encoding the strength of the fifth force is compatible with zero at all redshift and can be constrained with an error ranging from $0.17-0.29$, depending on redshift. Assuming a fifth force with constant strength over the range of observation, we constrain its amplitude to lie within $-21\%$ and $7\%$ of the gravitational interaction strength. Our method does not depend on specific theories for the fifth force. It relies however on the assumption that at high redshift the fifth force is negligible such that the matter power spectrum constrained by the CMB is recovered; and on the assumption that the background evolution of the Universe follows that of a $\Lambda$CDM model. These assumptions--which can in principle be relaxed--have indeed been applied in the redshift-space distortion analysis and the gravitational lensing analysis used for our constraints. We show that future surveys such as the Legacy Survey of Space and Time (LSST)~\cite{LSST2025} and DESI will improve the constraints, allowing us to detect a departure from pure gravitational interaction at the level of $3-6\%$ per redshift bin, over the range $z\in [0.51,1.35]$. Assuming a constant strength tightens the constraints to 2\%.

\section{Results} \label{sec:Results}

We combine redshift-space distortions with gravitational lensing measurement to constrain the validity of Euler's equation. As shown in~\cite{Bonvin:2018ckp}, dark matter interactions generically modify Euler's equation through two effects: an additional force encoded in the parameter $\Gamma(\eta)$ and a friction term encoded in the parameter $\theta(\eta)$:
\begin{align}
V'+(1+\theta)V-\frac{k}{\mathcal{H}}(1+\Gamma)\Psi=0\, .  
\label{eq:Euler}
\end{align}
In the cold dark matter scenario, $\Gamma=\theta=0$ applies.
Here, $V$ is the galaxy velocity potential in Fourier space, defined through $\bV(\mathbf{\bk},\eta)=i\bk/k V(\bk,\eta)$, a prime denotes derivatives with respect to the logarithm of the scale factor $a$, and $\mathcal{H}=(\mathrm da/\mathrm d\eta)/a$ is the Hubble parameter in conformal time $\eta$. In many models, the parameter $\theta$ is negligible compared to the parameter $\Gamma$ since it is sensitive to the time evolution of the field (scalar or vector) that governs dark matter interactions. In the quasi-static approximation the field evolves slowly and $\theta$ is negligible, see e.g.,~\cite{Khoury:2003aq,Hinterbichler:2010es,Bonvin:2018ckp}. Hence in the following we concentrate on the dominant effect and constrain $\Gamma$. 

Galaxy surveys and weak lensing surveys cannot measure directly the velocity field $\mathbf{V}(\bk,z)$ and the Weyl potential $\Psi_W(\bk,z)=\Psi(\bk,z)$ that enter into Euler's equation~\eqref{eq:Euler}. However, as discussed in Methods, the time evolution of these two fields can be measured from the galaxy correlation function and the galaxy-galaxy lensing correlation function. More precisely, redshift-space distortions provide direct measurements of the growth rate of structure $\hf$ that encodes the time evolution of the velocity field. Furthermore, combining galaxy-galaxy lensing with galaxy clustering provides direct measurements of the quantity $\hJ$ that encodes the time evolution of the Weyl potential. The amplitude of the fifth force $\Gamma$ can be expressed in terms of these two observable quantities as (see Section~\ref{sec:methods} for more detail)
\begin{align}
1+\Gamma(z)=\frac{2\hf(z)}{3\hJ(z)}\left(1-\frac{\mathrm d\ln \HH(z)}{\mathrm d\ln(1+z)}-\frac{\mathrm d\ln\hf(z)}{\mathrm d\ln(1+z)} \right)\, .   \label{eq:Gamma} 
\end{align}
Eq.~\eqref{eq:Gamma} is a key result of this paper. It shows that by combining measurements of $\hJ$, $\hf$, and its derivative at a given redshift $z$ we can directly measure the strength of the fifth force at that redshift.

\subsection{Constraints on Euler's equation with current data}
\label{sec:current}

\setlength{\tabcolsep}{0.8em} 
\begin{table}
\centering
\caption{\textbf{Mean values and 1$\sigma$ uncertainties of the Weyl evolution, the growth rate and the fifth force from current data} We list the first four effective redshifts of the DES \textsc{MagLim} sample along with the respective values of $\hat{J}(z)$ obtained in Ref.~\cite{Tutusaus:2023aux} (using CMB priors and standard scale cuts), and the values of $\hat{f}(z)$, $\mathrm d\ln\hat{f}(z)/{\mathrm d\ln(1+z)}$ and $\Gamma(z)$, with their 1$\sigma$ uncertainty, obtained in this work.} \label{tab:Jhat_fhat}
\begin{tabular}{@{}c c c c c@{}} 
\toprule
$z$ & $\hat{J}(z)$ & $\hat{f}(z)$ & $\frac{\mathrm d\ln\hat{f}(z)}{\mathrm d\ln(1+z)}$ & $\Gamma(z)$ \\
\midrule
\multicolumn{4}{c}{\vspace{-8pt}} \\
$0.295$ & $0.325 \pm 0.015$ & $0.459 \pm 0.019$ & $\phantom{-}0.28\pm 0.17$ & $-0.09\pm 0.17$ \\
$0.467$ & $0.333\pm 0.018$ & $0.467 \pm 0.020$ & $\phantom{-}0.00 \pm 0.17$ & $\phantom{-}0.04 \pm 0.17$\\
$0.626$ &  $0.387\pm 0.027$ & $0.461 \pm 0.021$ & $-0.24\pm 0.24$ & $-0.01 \pm 0.20$ \\
$0.771$ & $0.354\pm 0.035$  & $0.448 \pm 0.024$ & $-0.44\pm 0.31$ & $\phantom{-}0.16 \pm 0.29$ \\
\bottomrule
\end{tabular} 
\end{table}

\begin{figure}[h]
    \centering
    \includegraphics[height = 6.7cm]{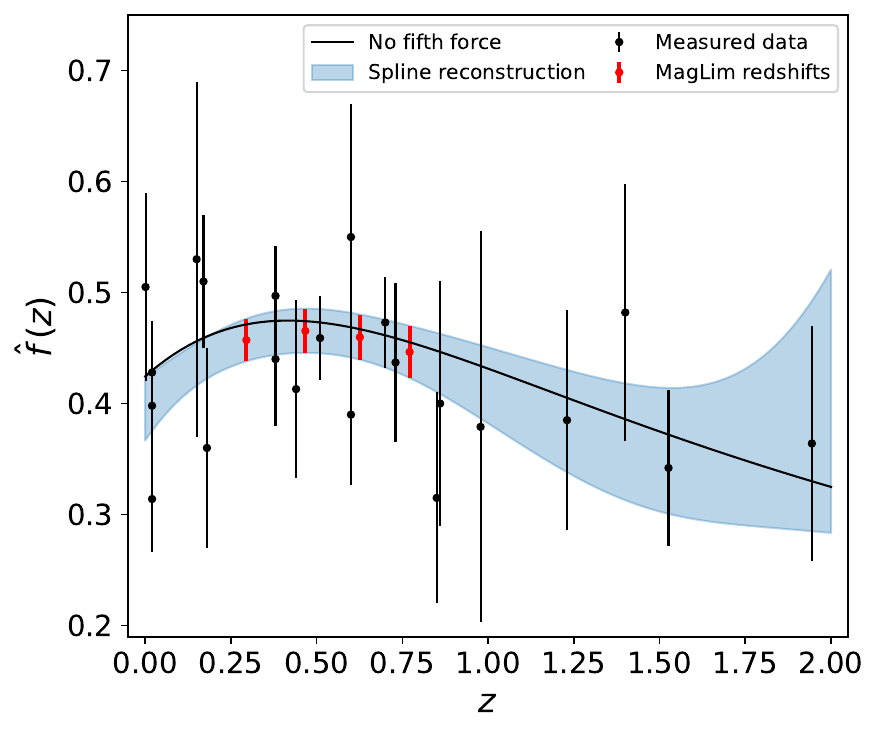}
    \includegraphics[height = 6.7cm]{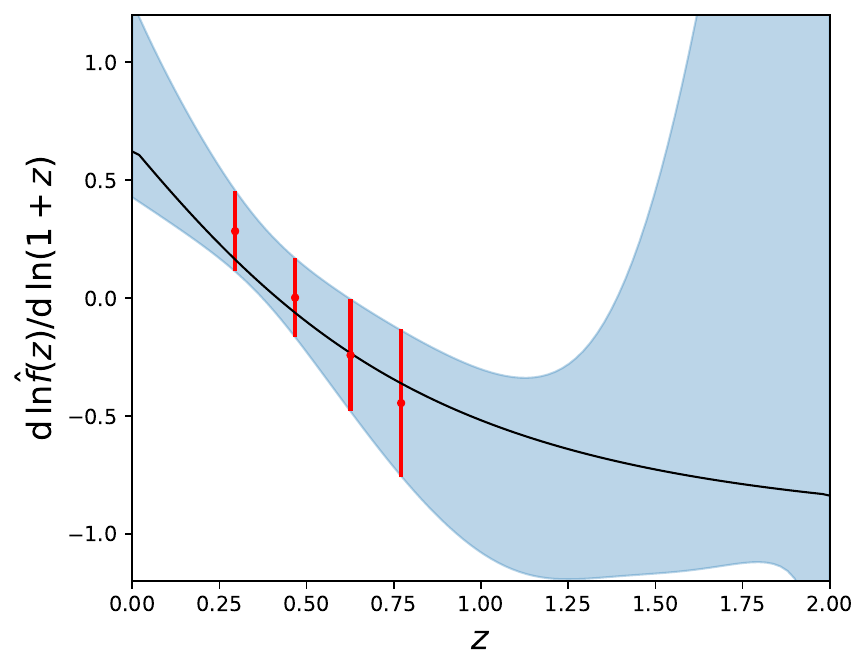}
    \caption{\textbf{Reconstruction of $\hf$ and its derivative with current data} \textit{Left panel:} The 22 measured data points of $\hat{f}$, from Table~1 of Ref.~\cite{Grimm:2024fui} (black points) and their spline reconstruction with 1$\sigma$ uncertainty (blue band), leading to the values of $\hat{f}$ at the four \textsc{MagLim} effective redshifts (red points). \textit{Right panel:} Reconstruction of $\mathrm d \ln \hat{f}(z)/\mathrm d\ln(1+z)$ based on the spline interpolation of $\hat{f}$. For both panels, the prediction assuming no fifth force and cosmological parameters from Planck~\cite{Planck:2018vyg} is shown as well (black line), being in agreement with the reconstruction at the $1\sigma$ level. Source data are provided as a Source Data file~\cite{codeFifthForce}.}\label{fig:f_current}
\end{figure}

We use 22 measurements of $\hf$ between redshifts $z = 0.001$ and $z = 1.944$, from various spectroscopic galaxy surveys~\cite{Howlett:2017asq,Huterer:2016uyq,Hudson:2012gt, Turnbull:2011ty,Davis:2010sw,Song:2008qt,Blake:2013nif,eBOSS:2020yzd,Blake:2012pj,Pezzotta:2016gbo,Okumura:2015lvp,eBOSS:2018yfg}. The measurements with their uncertainties are listed in Table I of~\cite{Grimm:2024fui}. Note that we do not include the new measurements from DESI~\cite{DESI:2024jis}, since we do not have the covariance of these measurements with the other 22 bins. We have checked that adding these measurements would reduce the uncertainty on $\hf$ over the range where we measure $\Gamma$ by at most $20\%$. Our aim is to infer $\Gamma(z)$ at the four DES \textsc{MagLim} effective redshifts where we have measurements of $\hJ$: $z\in \{
0.295, 0.467, 0.626, 0.771\}$. We use therefore the 22 measurements of $\hf$ to reconstruct $\hf$ and its redshift derivative $\mathrm d\ln\hf(z)/\mathrm d\ln(1+z)$ at those redshifts. We treat the four values of $\hf$ at the DES effective redshifts as free parameters and we interpolate between these parameters using cubic spline interpolation.
We then determine the values of $\hf$ at the effective redshifts by minimizing the difference between the interpolated curve and the measurements of $\hf$. The reconstructed values of $\hf$ at the four \textsc{MagLim} effective redshifts, together with the reconstruction over the whole redshift range, are plotted in the left panel of Fig.~\ref{fig:f_current} and listed in Table~\ref{tab:Jhat_fhat}. The spline reconstruction also allows us to infer the redshift derivatives $\mathrm d\ln\hf(z)/\mathrm d\ln(1+z)$ and their uncertainty at the desired redshifts. The results are plotted in the right panel of Fig.~\ref{fig:f_current} and listed in Table~\ref{tab:Jhat_fhat}. Note that instead of treating the values of $\hf$ at the four DES redshifts as free parameters, we could instead choose a different set of redshift knots and treat the values of $\hf$ at those knots as free parameters that we determine through interpolation and optimization. The optimal number of redshift knots is then the one that minimises the Akaike information criterion (AIC)~\cite{Akaike:1974vps}, ensuring a good fit of the data, while preventing overfitting. For the data set used in this analysis, we find that four knots is the optimal choice, and we comment more on the reconstruction with different numbers of knots in Supplementary Discussion~\hyperref[Appendix:Numbers_knots]{1}. Moreover, we find that the reconstruction of $\hf$ does not depend on the placement of these four knots, since a cubic spline interpolation with four knots and the standard \textit{not-a-knot} boundary condition reduces to a fit with a single third-degree polynomial.
Therefore, we find the same values for $\hf$ and its derivative at the DES redshifts when using this method.

\begin{figure}[h]
    \centering
    \includegraphics[width=.495\textwidth]{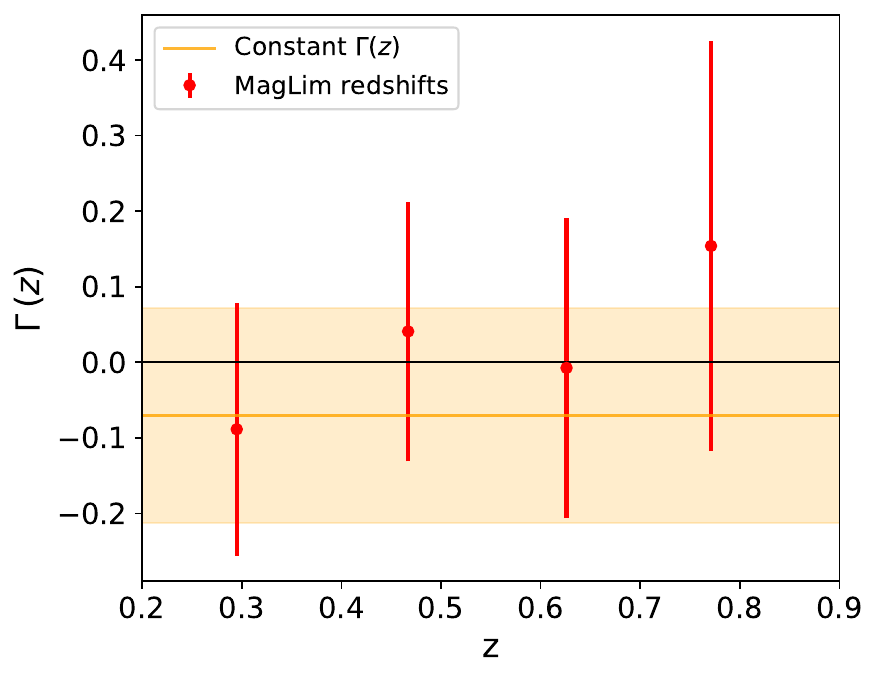}
    \caption{\textbf{Constraints on $\Gamma$ with current data} We show the reconstructed values (in red) of the fifth force parameter $\Gamma$ together with the 1$\sigma$ uncertainties at the four effective redshifts of the DES \textsc{MagLim} sample. The measurements show no deviation from Euler's equation ($\Gamma=0$, black horizontal line). The green line with error bands shows the best-fit value and $1\sigma$ uncertainty assuming a constant value of $\Gamma$.  We note that the measurements at different redshifts are correlated, as can be seen from the covariance matrix given in Supplementary Discussion~\suppref{app:covariance}{2}. Source data are provided as a Source Data file~\cite{codeFifthForce}. }\label{fig:current}
\end{figure}

We then use the values of the $\hf$ derivatives, together with the values of $\hJ$ listed in Table~\ref{tab:Jhat_fhat}, to constrain $\Gamma$ following Eq.~\eqref{eq:Gamma}. The results for $\Gamma$ are plotted in Fig.~\ref{fig:current} and listed in the last column of Table~\ref{tab:Jhat_fhat} (see also the covariance matrix listed in Supplementary Discussion~\suppref{app:covariance}{2}). We see that $\Gamma$ is compatible with zero at all redshifts: current data show therefore no violation of Euler's equation for dark matter. Moreover, our results put constraints on the allowed amplitude of the fifth force in each redshift bin. From Euler's equation~\eqref{eq:Euler}, we see that gravitational interaction affects the motion of galaxies through the term $(k/\mathcal{H})\Psi$, while the impact of the fifth force is given by $\Gamma \times (k/\mathcal{H})\Psi$. This allows us to compare the strength of the fifth force with that of gravitational interaction. For example, in the first redshift bin, the fifth force is constrained to be within $-26\%$ and $8\%$ of the gravitational interaction strength.

We then assume a fifth force with a constant amplitude within the observed redshift range, and combine the four measurements (accounting for their covariance). We find that in this case $\Gamma=-0.07\pm 0.14$, meaning that the amplitude of the fifth force is constrained to be within $-21\%$ and $7\%$ of the gravitational interaction strength, see green band in Fig.~\ref{fig:current}. If instead we assume that $\Gamma$ increases proportionally to dark energy, which could be the case if dark matter interacts with dark energy, we find that the value of $\Gamma$ today is constrained to $\Gamma(z=0)=-0.12\pm 0.22$, leading to the following values at the DES redshifts: $\Gamma\in$ $\{-0.09\pm 0.16$, $-0.07\pm0.13$, $-0.06\pm 0.11$, $-0.05\pm 0.09\}$. Finally, we examine the case where $\Gamma$ is restricted to be strictly positive. This is motivated by specific models of dark matter, for example the coupling quintessence models explored in~\cite{Castello:2024jmq}, where $\Gamma$ depends on the square of the coupling strength, thus not allowing any negative fifth force. Under this restriction, we find that the fifth force cannot exceed $11\%$ of the gravitational interaction, i.e.~$\Gamma \leq 0.11$ for the 68\% upper limit (and $\Gamma \leq 0.24$ for the 95\% upper limit).

These constraints can be compared with the recent constraints on a dark fifth force, in a specific model with a long-range dark force mediated by an ultralight scalar field, derived in~\cite{Bottaro:2023wkd}: using CMB alone, the strength of the fifth force is constrained to be below 1.2\%. Adding Baryon Acoustic Oscillations (BAO) tightens the constraints to 0.48\%. These constraints are however not directly linked to a breaking of Euler's equation, but rather driven by the background evolution of dark matter density in the model considered. More precisely, due to the additional coupling, the dark matter energy density does not decay anymore as $1/a^3$. This directly changes the redshift-distance relation in the Universe, that differs from $\Lambda$CDM predictions, leading to very tight constraints on the coupling. In practice, however, we do not know what is causing the accelerated expansion of the Universe. It could be a cosmological constant, or it could be a dynamical scalar field. Changes in distances induced by a dark fifth force are fully degenerated with changes induced by a dynamical dark energy, with equation of state parameter $w\neq -1$. In our work, we explore therefore a fully different scenario: since a signature at the level of the background cannot uniquely point to the presence of a dark fifth force, we consider that the impact of both the fifth force and any ingredient impacting the background evolution (for example the quintessence field in the case of coupled quintessence) can be encoded into an effective equation of state $w_{\rm eff}$. This is the approach followed, e.g.,\ in~\cite{Wang:2023tjj,Castello:2024jmq}. This $w_{\rm eff}$ is constrained to be close to $-1$ by distance measurements and we therefore fix it to this value in our analysis. We then constrain the fifth force by directly looking at its impact on Euler's equation, i.e., at the deviation it would induce on the way dark matter falls into a gravitational potential. The constraints that we obtain are an order of magnitude larger than those coming from the background evolution of dark matter density, but the advantage is that any deviations from $\Gamma=0$ would uniquely point to the presence of a dark fifth force. No (uncoupled) dynamical dark energy model can mimic or hide such a deviation. Our approach also has the advantage that it applies to interacting dark matter models with a pure momentum exchange, that do not alter the background, see e.g.~\cite{Pourtsidou:2013nha,Chamings:2019kcl}. 

Another key feature of our method is that it does not rely on any model for the fifth force evolution. We do not need to specify the form of the dark matter interaction, nor the characteristic of the field propagating the fifth force, such as the form of its potential, or its nature (scalar or vector). Our constraints can therefore be used to constrain any model of interest, without redoing the analysis, since $\Gamma$ can be related to the parameters of the model, see e.g.,~\cite{Bonvin:2018ckp}.

\subsection{Forecasts with future surveys}

\setlength{\tabcolsep}{0.8em} 
\begin{table}
\centering
\caption{\textbf{Mean values and 1$\sigma$ uncertainties of the Weyl evolution, the growth rate and the fifth force from future surveys} We list the first nine effective redshifts of the LSST sample along with the respective values of $\hat{J}(z)$ forecasted in Ref.~\cite{Tutusaus:2022cab} with 1$\sigma$ uncertainties (using the pessimistic case), the values of $\hat{f}(z)$ and $\mathrm d\ln\hat{f}(z)/{\mathrm d\ln(1+z)}$ obtained at the same redshifts using the DESI forecasts~\cite{DESI:2016fyo} and spline interpolation, as well as the resulting values of $\Gamma(z)$.} \label{tab:Jhat_fhat_future}

\begin{tabular}{@{}c c c c c @{}} 
\toprule
$z$ & $\hat{J}(z)$ & $\hat{f}(z)$ & $\frac{\mathrm d\ln\hat{f}(z)}{\mathrm d\ln(1+z)}$ & $\Gamma(z)$ \\
\midrule
\multicolumn{4}{c}{\vspace{-8pt}} \\
$0.25$ & $0.333\pm 0.002$ & $0.468 \pm 0.009$ & $0.23 \pm 0.19$ & $0.00\pm 0.17$  \\
$0.38$ & $0.360\pm 0.002$ & $0.474\pm 0.007$ & $0.05\pm 0.10$ & $-0.01\pm 0.09$ \\
$0.51$ & $0.378\pm 0.003$ & $0.473\pm 0.006$  & $-0.11\pm 0.05$ & $0.00 \pm 0.05$ \\
$0.65$ & $0.388\pm 0.003$ & $0.466\pm 0.004$  & $-0.26\pm 0.07$ & $0.00 \pm 0.06$ \\
$0.79$ & $0.391\pm 0.003$ & $0.453 \pm 0.003$  & $-0.39\pm 0.07$ & $0.01 \pm 0.06$ \\
$0.95$ & $0.388\pm 0.004$ & $0.437\pm 0.002$  &  $-0.48 \pm 0.04$ & $0.00 \pm 0.03$ \\
$1.13$ & $0.380\pm 0.004$  & $0.417\pm 0.002$   & $-0.58\pm 0.06$ & $0.00 \pm 0.05$\\
$1.35$ & $0.337\pm 0.004$ & $0.392\pm 0.003$ & $-0.69\pm 0.07$ & $0.01 \pm 0.05$ \\
$1.70$ & $0.306\pm 0.004$ & $0.354\pm 0.008$ & $-0.76\pm 0.39$ & $-0.01\pm 0.26$ \\
\bottomrule
\end{tabular} 
\end{table}

\begin{figure}[h]
    \centering
    \includegraphics[height = 6.7cm]{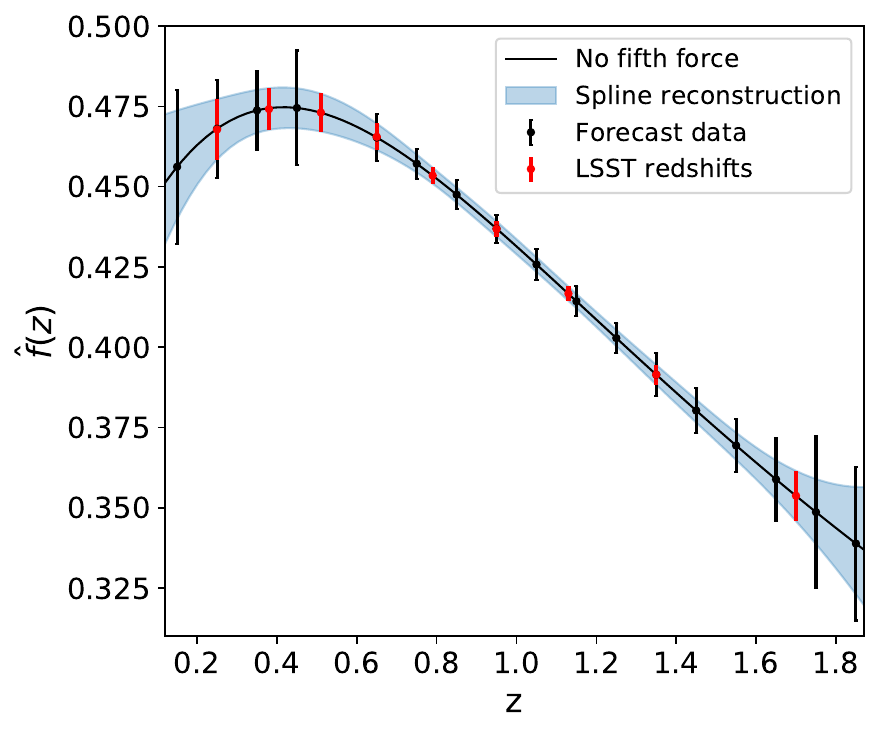}
    \includegraphics[height = 6.7cm]{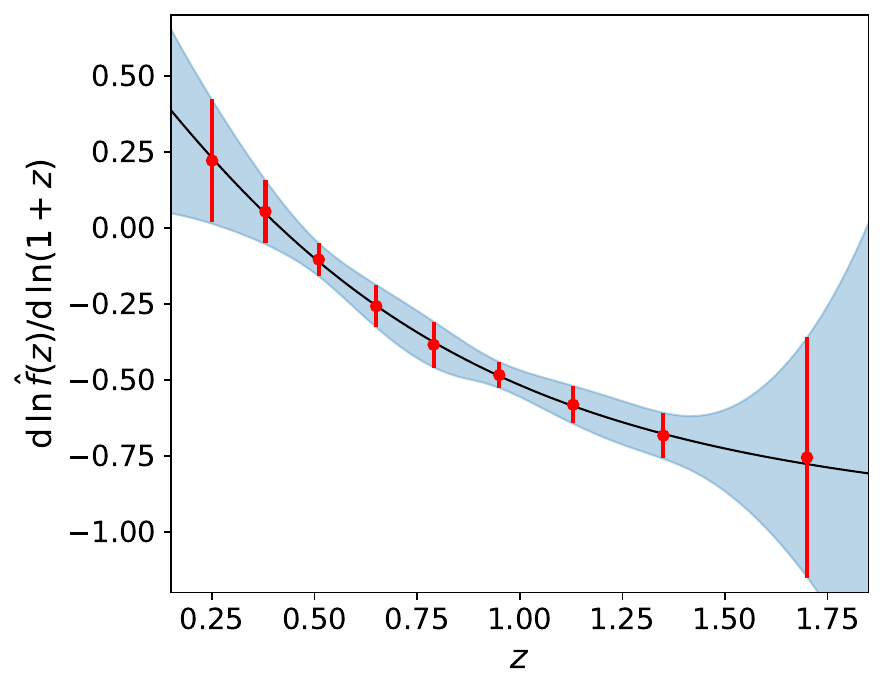}
    \caption{\textbf{Reconstruction of $\hf$ and its derivative with future surveys} \textit{Left panel:} We show 17 values for $\hat{f}$ centered around the $\Lambda$CDM fiducial (black points) and with 1$\sigma$ uncertainties achievable by DESI covering 14,000 square degrees (see Tables~2.3 and~2.5 of~\cite{DESI:2016fyo}). We also show the spline reconstruction (blue band), leading to the values of $\hat{f}$ at the nine effective redshifts of LSST (red points). \textit{Right panel:} Reconstruction of $\mathrm d \ln\hat{f}(z)/\mathrm d\ln(1+z)$ based on the spline interpolation of $\hat{f}$. For both panels, the prediction without a fifth force is shown as well (black line). Source data are provided as a Source Data file~\cite{codeFifthForce}.
    } \label{fig:f_future}
\end{figure}

\begin{figure}[h]
    \centering
    \includegraphics[width=.495\textwidth]{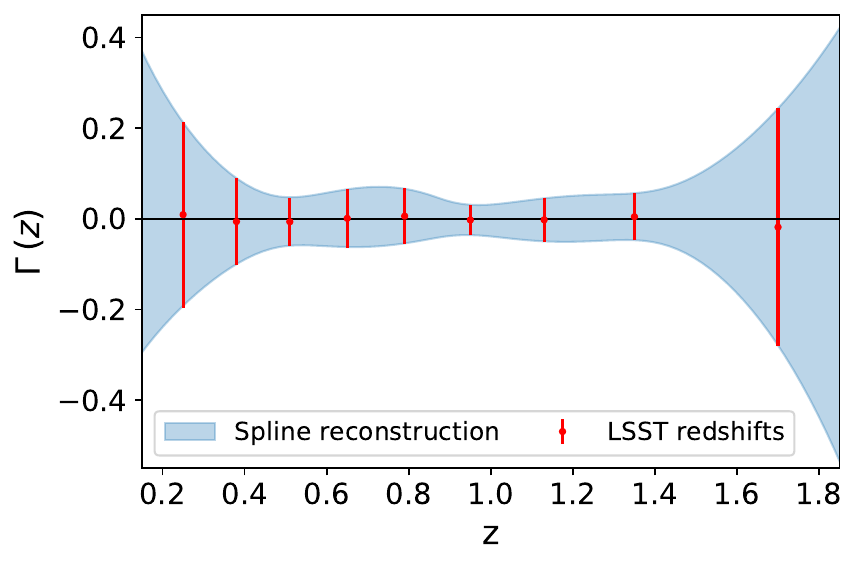}
    \caption{\textbf{Constraints on $\Gamma$ with future surveys} Using forecast values of $\hat{f}$ from DESI~\cite{DESI:2016fyo}, we show the reconstructed values (in red) of the fifth force parameter $\Gamma$ together with the 1$\sigma$ uncertainties at the nine effective redshifts corresponding to the LSST forecast for $\hat{J}$~\cite{Tutusaus:2022cab}. Additionally, the blue band shows a forecast over the whole redshift range when interpolating the $\hat{J}$ data as well. Source data are provided as a Source Data file~\cite{codeFifthForce}.}
    \label{fig:future}
\end{figure}

Our constraints rely on 22 measurements of $\hf$ from past and current spectroscopic surveys and 4 measurements of the Weyl evolution $\hJ$ from photometric DES data. The coming generation of surveys, including DESI, Euclid, LSST, and SKAO, holds the potential to drastically improve on these measurements. We forecast therefore the uncertainty on the fifth force $\Gamma$, from a combination of forecast values of $\hf$ from DESI and of $\hJ$ from LSST. More precisely, we use values of $\hf$ at 17 redshifts between $z=0.15$ and $z=1.85$ and with $1\sigma$ uncertainties as specified in Table~2.3 and Table~2.5 of Ref.~\cite{DESI:2016fyo} (we omit the lowest redshift value at $z=0.05$, as it has a larger uncertainty and no impact on our results). These specifications assume that DESI realises its full 14,000 square degrees of survey area, and obtains spectroscopic redshifts of more than 30 million galaxies. For LSST, we use the pessimistic uncertainties forecasted in~\cite{Tutusaus:2022cab} at nine redshifts between $z=0.25$ and $z=1.7$ (omitting the forecast at $z=2.1$ since this is well beyond the range where $\hat{f}$ data from DESI will be available), see the first two columns of Table~\ref{tab:Jhat_fhat_future}. The pessimistic uncertainties for $\hJ$ are more conservative and do not degrade the results for $\Gamma$, since those are dominated by the uncertainties on the derivative of $\hat{f}$. We center the values of $\hf$ and $\hJ$ around their prediction using cosmological parameters from Planck~\cite{Planck:2018vyg} and assuming no fifth force. For $\hJ$, we account for the covariance between redshift bins, which is non-zero due to the non-negligible overlap of the photometric redshift bins (see Fig.~1 of~\cite{Tutusaus:2022cab}). For DESI, we neglect the covariance between the bins, which is expected to be small due to the sharp edges of the bins in spectroscopic samples. This assumption can be tested once data from the completed DESI survey are available, and if needed the covariance can easily be included. 

On the left panel of Fig.~\ref{fig:f_future}, we plot the forecast data for $\hat{f}$ as well as their interpolation over the whole redshift range, $z\in [0.15, 1.85]$, again using spline interpolation between a number of redshift knots. We find that the optimal choice is five knots with the central one located at $z=0.87$. The AIC for four and five knots is actually very similar (even a little larger for five knots), but as we show in Supplementary Discussion~\suppref{Appendix:Numbers_knots}{1}, four knots lead to a slightly worse reconstruction of $\Gamma$ (at the 1$\sigma$ level). Hence, we adopt five knots as our baseline case. As an alternative interpolation method, we have as well considered Gaussian processes, as recent research has suggested that they may be applicable to next-generation large-scale structure data~\cite{Perenon:2021uom}. However, for the DESI specifications applied in this work, we have found that Gaussian processes lead to a biased reconstruction of $\hat{f}$ and particularly its derivative, showing a deviation from the fiducial model. Thus, we have chosen to show results for spline interpolation only.

In the right panel of Fig.~\ref{fig:f_future}, we show the resulting reconstruction of $\mathrm d\ln \hat{f}(z)/\mathrm d\ln(1+z)$. The values of these quantities at the LSST effective redshifts, where future measurements of $\hat{J}$ will be available, are indicated in red in the figure and listed in the third and fourth column of Table~\ref{tab:Jhat_fhat_future}. 
Finally, in Fig.~\ref{fig:future}, we show (in red) the results for $\Gamma$ at the LSST redshifts. As we have nine values of $\hat{J}$ with very high precision, we can perform a spline interpolation between these values, and therefore obtain a reconstruction (in blue) of $\Gamma$ along the whole redshift range, $z\in [0.15, 1.85]$. The constraints at the LSST redshifts are also listed in the last column of Table~\ref{tab:Jhat_fhat_future}.  We see that the constraints are significantly tighter than current ones. We also note that the mean values for $\Gamma$ are, as a result of the spline interpolation, not always exactly equal to the fiducial value of zero. However, the fiducial value is always well contained within the $1\sigma$ error bars, meaning that the interpolation method does not lead to any false imprints of new physics. In particular, over the range $z\in [0.51,1.35]$, we find that DESI combined with LSST will allow to detect a departure from pure gravitational interaction at the level of $3-6\%$ per redshift bin. Assuming a constant strength of the fifth force and applying the results for $\Gamma$ at the LSST redshifts (including their covariance), we find that the combination of DESI and LSST data will allow to constrain a fifth force with amplitude down to 2\% of the gravitational interaction strength.

In our analysis and forecasts, we have tested Euler's equation, under the assumption that general relativity is valid at cosmological scales. In this scenario, any deviation in Euler's equation would be due to non-gravitational dark matter interactions. If general relativity is not valid however, Euler's equation could be violated by the new degree of freedom mediating gravity, that could break the equivalence principle between standard matter and dark matter, see, e.g.,~\cite{Gleyzes:2015pma}. Alternatively, in models beyond general relativity, the time distortion and the spatial distortion can be different~\cite{Amendola:2007rr,Daniel:2008et}, leading to an apparent breaking of Euler's equation due to the fact that we used the Weyl potential $\Psi_W$ instead of the time distortion $\Psi$. To distinguish between these scenarios, one would need to measure directly the time distortion $\Psi$. As shown in~\cite{Sobral-Blanco:2022oel}, this will be possible with future surveys like DESI, Euclid, and the SKAO, by looking at the impact of gravitational redshift on the distribution of galaxies. Combining these new measurements of $\Psi$ with that of the Weyl potential and of the galaxy velocities will allow us to distinguish between a non-gravitational interaction that would affect only Euler's equation, a modification of gravity that would generate a difference between $\Psi$ and $\Psi_W$~\cite{Bonvin:2022tii,Castello:2024jmq}, and a modified gravity model that would break the weak equivalence principle and modify both Euler's equation and the relation between $\Psi$ and $\Psi_W$. 

\section{Discussion}

In this paper we have performed a direct test of the validity of Euler's equation for dark matter at cosmological scales. We have combined measurements of galaxy peculiar velocities with measurements of the Weyl potential, to place constraints on the existence of a fifth force that would alter the way dark matter falls inside a gravitational potential. We have found that current data do not favor the existence of such a fifth force, and we have placed constraints on the strength of interaction in four redshift bins. Moreover, assuming that the strength of the fifth force is constant over our range of observation, we have found that a positive fifth force cannot exceed 7\% of the gravitational interaction strength, while a negative fifth force is constrained to be less than 21\% of the strength. Future data will improve the precision and allow to detect departures from pure gravitational interaction at the level of $3-6\%$ per redshift bin, and at 2\% assuming a constant amplitude.

In our analysis we have let the parameter encoding the strength of the fifth force take any sign. In specific models of dark matter interactions, the sign is determined by the physical impact of the interaction. Of course, not all interacting dark matter models can be described by Eq.~\eqref{eq:Euler} with a free parameter $\Gamma$. However, generally, even more complicated interactions can be absorbed in an effective $\Gamma$, that may not have a physical meaning (and can depend on ratio of perturbations) but that effectively captures a deviation in Euler's equation.  In practice, in some models of dark energy coupled with dark matter, the fifth force $\Gamma$ is always positive (proportional to the square of the coupling strength) and enhances the clustering of dark matter~\cite{Castello:2024jmq}. On the other hand, if dark matter interacts with dark radiation, it can lead to a force that effectively reduces the clustering with respect to pure gravitational interaction~\cite{Archidiacono:2019wdp}, which can be represented by a negative $\Gamma$. Similarly, specific models of dark matter coupled to dark energy with a pure momentum exchange also lead to an effective decrease of dark matter clustering~\cite{Chamings:2019kcl}. Finally, in the case where gravity is modified and the weak equivalence principle is broken, dark matter can feel a larger or smaller interaction than baryons, leading to any sign for $\Gamma$.

The presence of a fifth force acting on dark matter would not only break Euler's equation, but it would also leave an impact on the evolution of the density fluctuations and the gravitational potentials. As shown in~\cite{Castello:2022uuu}, the impact of $\Gamma$ on the density evolution is exactly the same as the impact of a modification to Poisson's equation (generated by gravity modifications), encoded in the so-called parameter $\mu$~\cite{Amendola:2007rr,Bertschinger:2008zb,2010PhRvD..81j4023P}. Current constraints on $\mu$ from redshift-space distortions can therefore directly be translated into constraints on $\Gamma$. From the recent analysis of DESI~\cite{DESI:2024hhd}, we see that those constraints are at the level of 45\% (assuming that $\mu$ evolves proportionally to dark energy). Adding CMB constraints from Planck and gravitational lensing from DES tightens the constraints on $\mu$ to 22\%. 
Since $\Gamma$ is expected to be constrained at a similar level from its impact on the growth of structure, we see that our constraints, derived from Euler's equation, are actually better than indirect constraints. 
In addition to impacting the growth of structure, a dark fifth force would also source relative density and velocity fluctuations between dark matter and baryons that affect BAO. However, as shown in~\cite{Bottaro:2023wkd}, this effect is subdominant with respect to the impact of the fifth force on the growth of structure. It therefore does not tighten current constraints on $\Gamma$. Hence we conclude that our constraints from Euler's equation are stronger than indirect constraints, while having the advantage of not being degenerated with $\mu$, since modifying Poisson's equation has no impact on Euler's equation. 

A remarkable characteristic of our analysis is that it does not require to specify the type of dark matter interaction responsible for the breaking of Euler's equation, nor its time evolution. The only assumption is that at early time, the fifth force is negligible such that one recovers the matter power spectrum constrained by CMB. Since galaxy velocities as well as the Weyl potential can be measured at different moments of the history of the Universe, they can be used to test Euler's equation redshift bin by redshift bin. This is particularly interesting in the case where dark matter would interact with dark energy, whose impact becomes more and more relevant at low redshift. In such a scenario, one could expect a fifth force growing with time. Future surveys, that will provide measurements of the growth rate $\hat{f}$ and the Weyl potential $\hat{J}$ in a larger number of bins, and over a larger redshift range, will make optimal use of this characteristic, allowing a refined reconstruction of the evolution of the fifth force. Finally, while not being a subject of this work, we note that the high precision of future surveys may allow us to probe a scale-dependence of $\hat{f}$ as well as $\hat{J}$. Our method to constrain the fifth force by combining these quantities could be easily extended to such cases, taking the binning in scale in addition to the binning in redshift into account. Thus, future surveys hold the potential to provide precise results on the existence of a fifth force, as well as on its scale- and redshift-behaviour.

\section{Methods} \label{sec:methods}

We assume a perturbed Friedmann-Lema\^itre-Robertson-Walker universe, with a homogeneous and isotropic background plus perturbations, whose geometry is encoded in the metric:
\begin{align}
\mathrm ds^2=a^2(\eta)\left[-(1+2\Psi(\mathbf{x},\eta))\mathrm d\eta^2+ (1-2\Phi(\mathbf{x},\eta))\delta_{ij}\mathrm dx^j\mathrm dx^j\right]\,.
\end{align}
Here $a$ is the scale factor, $\eta$ denotes conformal time, and the two gravitational potentials $\Psi$ (time distortion) and $\Phi$ (spatial distortion) encode the perturbations of the geometry. In addition, the perturbations in the matter content can be encoded into two extra fields, namely the fluctuations in the galaxy density $\delta_{\rm g}=\delta\rho_{\rm g}/\rho_{\rm g}$ and the galaxy peculiar velocity $\bV_{\rm g}$. In the following we assume that the galaxy velocity is governed by the velocity of the dark matter halo $\bV_{\rm g}=\bV_{\rm dm}$ and we drop the subscript $\mathrm{dm}$. 

The aim is to use cosmological data to constrain the strength of a possible fifth force, denoted by $\Gamma$, acting on dark matter. From Eq.~\eqref{eq:Euler} and neglecting the subdominant friction term $\theta$, we obtain
\begin{align}
\label{eq:Gamma_V}
1+\Gamma=\frac{\HH}{k}\frac{(V'+V)}{\Psi}\, .    
\end{align}
Hence, $\Gamma$ can be directly constrained from measurements of $V$ and $\Psi$.
Note that if instead of assuming that the velocity of galaxies is fully driven by the velocity of dark matter halos, we account for a fraction of baryons that obey Euler's equation, the galaxy velocity becomes a weighted average of the dark matter velocity and of the baryon velocity (denoted by $V_{\rm b}$): $V_{\rm g}=xV_{\rm dm}+(1-x)V_{\rm b}$, where $x=\rho_{\rm dm}/(\rho_{\rm dm}+\rho_{\rm b})\simeq 0.8$. In this case, as shown in Supplementary Discussion~\hyperref[app:velocity]{3}, $\Gamma$ is replaced by $x\Gamma$ in Eq.~\eqref{eq:Gamma_V}. This degrades the constraints on $\Gamma$ by a factor $x$.

Galaxy surveys and weak lensing surveys cannot measure directly the velocity potential $V(\bk,z)$ and the Weyl potential $\Psi_W(\bk,z)=\Psi(\bk,z)$ that enter into Eq.~\eqref{eq:Gamma_V}. However, the time evolution of these two fields can be measured and used to constrain $\Gamma$. More precisely, the velocity potential at redshift $z$ can be written in terms of an initial velocity at $z_*$ as 
\begin{align}
V(k,z)=\frac{\HH(z)\hf(z)}{\HH(z_*)\hf(z_*)}V(k,z_*)\, ,
\label{eq:hatf}
\end{align}
where $\hf(z)=f(z)\sigma_8(z)$, with $f(z)=\mathrm d\ln \delta/\mathrm d\ln a$ is the growth rate of structure 
and $\sigma_8(z)$ the amplitude of density perturbations in spheres of 8\,$h^{-1}$Mpc. The function $\hf(z)$ directly encodes the evolution of velocities and it can be measured from spectroscopic redshift surveys like the Sloan Digital Sky Survey (SDSS)~\cite{SDSS2025} and the WiggleZ Dark Energy Survey~\cite{WiggleZ2025}.
These surveys measure indeed the multipoles of the two-point correlation function (or power spectrum in Fourier space). 
Using Eq.~\eqref{eq:hatf}, these multipoles can be expressed in terms of three quantities only: the matter density power spectrum at early time $P_{\delta\delta}(k,z_*)$, the growth rate $\hat{f}(z)$, and the galaxy bias $\hat{b}(z)=b(z)\sigma_8(z)$. One can then choose $z_*$ in the matter era, well before the accelerated expansion of the Universe started and use that, at that redshift, the density matter power spectrum is well constrained by measurements from the CMB. The multipoles provide then direct measurements of $\hat{b}$ and $\hat{f}$ at a set of redshift bins. The key point of this method is that it does not rely on a specific theory of gravity, a dark energy model or a dark matter model. Any possible deviation from the $\Lambda$CDM model between redshift $z_*$ and today is encoded in the function $\hat{f}(z)$. Hence, measurements of $\hf$ can be consistently used to test models beyond $\Lambda$CDM~\cite{eBOSS:2020yzd}.

There are however two assumptions in such measurements of $\hf$. The first one is that at $z_*$, we recover the matter power spectrum of a cold dark matter universe, see e.g.,~\cite{BOSS:2016ntk}. This is consistent with measurements from the CMB that place tight constraints on $P_{\delta\delta}(k,z_*)$. This means that in our analysis we have to limit ourselves to models where the fifth force is negligible at early time. This is typically the case if the fifth force is due to interactions of dark matter with dark energy, which is fully negligible at early time. If the fifth force is, however, due to self-interacting dark matter or dark matter interacting with dark photons, then its evolution with time is a priori unknown and it depends on the particular model. In this case, one would either need to modify the matter power spectrum at $z_*$ and redo the measurements of the growth rate $\hf(z)$ for each model. Or we could leave the matter power spectrum free and constrain it in $k$-bands together with the growth rate, as proposed in~\cite{Schirra:2024rjq,Castello:2024lhl}. The second assumption in current measurements of $\hf$ is that it does not depend on $k$. This is strictly correct for cold dark matter and within general relativity (the scale-dependence due to massive neutrinos is indeed negligible for a neutrino mass sum of 0.06 eV~\cite{Euclid:2019clj}, which is what we assume in our analysis). Adding a fifth force may introduce a scale-dependence of $\hat{f}$, however data are currently not constraining enough to test a scale-dependence~\cite{BOSS:2016ntk,Amendola:2022vte}, and we can therefore omit it. Moreover, in various models of dark matter interactions, for example interactions mediated by a scalar field or a vector field, the fifth force is actually scale-independent at sub-horizon scales~\cite{Bonvin:2018ckp}.

In addition to $\hf$, to test Euler's equation, we also need measurements of the Weyl potential. The Weyl potential governs the trajectory of light emitted by distant galaxies, and generates distortions in their observed shape. These distortions have been measured by various surveys and then used to infer the distribution of matter in the Universe, see, e.g.,~\cite{KiDS:2020suj,DES:2021wwk}. Recently, Ref.~\cite{Tutusaus:2022cab} designed a novel method that uses weak lensing data to directly measure the evolution of the Weyl potential across redshift. The idea is very similar to that used for $\hf$ measurements: we assume a known matter power spectrum at redshift $z_*$, well constrained by CMB, and we encode the evolution of the Weyl potential into a free function of redshift. No assumptions are made about the evolution of this function. More precisely, we write the Weyl potential as
\begin{align}
\Psi_W(k,z)= \left(\frac{\HH(z)}{\HH(z_*)} \right)^2 \sqrt{\frac{B(k,z)}{B(k,z_*)}}\hJ(z)\frac{\Psi_W(k,z_*)}{\sigma_8(z_*)}\, ,
\label{eq:hatJ}
\end{align}
where $\hJ$ encodes the evolution of $\Psi_W(k,z)$ and $B(k,z)$ is a boost factor, accounting for the non-linear evolution of matter density perturbations at small scales. Note that since the Weyl potential is related to the matter density perturbations through Einstein's equations, the function $\hJ$ is proportional to $\sigma_8(z)$ within general relativity: $\hJ(z)=\Omega_{\rm m}(z)\sigma_8(z)$, where $\Omega_{\rm m}$ is the matter density parameter.  As shown in~\cite{Tutusaus:2022cab,Tutusaus:2023aux}, the galaxy-galaxy lensing correlation function and the galaxy clustering correlation function can be used to measure $\hJ$. They can indeed be written in terms of four quantities: the matter density power spectrum at early time $P_{\delta\delta}(k,z_*)$, the Weyl evolution $\hJ(z)$, the galaxy bias $\hat{b}(z)$, and the boost factor $B(k,z)$. As before, $P_{\delta\delta}(k,z_*)$ is assumed to be that of a cold dark matter universe. The cosmological parameters affecting $P_{\delta\delta}(k,z_*)$ and the functions $\hat{b}$ and $\hJ$ are then measured together from the data. Since the CMB provides tight constraints on early universe physics, we add priors on the cosmological parameters when we vary them, corresponding to the $3\sigma$ constraints obtained from Planck~\cite{Planck:2018vyg}. As for redshift-space distortions, this method has the advantage to provide measurements of $\hJ$ that do not depend on a specific dark matter model. They can therefore consistently be used to constrain $\Gamma$.

One non-trivial difference with respect to Eq.~\eqref{eq:hatf} is that $\Psi_W$ contains the boost factor $B(k,z)$. This boost is necessary to properly account for non-linearities that affect the lensing correlation function at small angular separation. Since we cannot write a boost factor in a model-independent way, we instead model it in general relativity, assuming cold dark matter, as done in~\cite{DES:2018ufa,DES:2021wwk,DES:2022ygi}. We also choose the same scale cuts as in~\cite{DES:2021wwk} to ensure that baryonic effects are mitigated. A fifth force acting on dark matter may however modify the growth of the Weyl potential at non-linear scales and thus impact the boost. This would however not invalidate our test: if $\Gamma=0$, then the boost is the correct one and our constraints are robust. If on the other hand $\Gamma\neq 0$, then the boost may be incorrect, meaning that the $\Gamma$ that we infer may not be exactly the one of the underlying dark matter model. However, the observation that $\Gamma\neq 0$, i.e.\ that a fifth force exists would still be robust: modelling the boost without a fifth force cannot wrongly lead us to find that a fifth force exists. It is only the interpretation of $\Gamma$ in terms of the underlying dark matter model that could be affected by an incorrect boost. If such a result is found, one could identify models that are consistent with the measured $\Gamma$, model the boost in such models and then redo the analysis.

We can now combine the measurements of $\hf$ and $\hJ$ to constrain $\Gamma$. Inserting Eqs.~\eqref{eq:hatf} and~\eqref{eq:hatJ} into~\eqref{eq:Gamma_V} we obtain Eq.~\eqref{eq:Gamma}, where we have used Einstein's equations and the continuity equation to relate $\Psi_W(k,z_*)$ and $V(k,z_*)$ to $\delta(k,z_*)$. Note that even though $\hJ$ and $\hf$ were not measured at the same scales (lensing measurements probe indeed smaller scales than redshift-space distortions), we can use them to test Euler's equation in the linear regime. Indeed, since both $\hJ$ and $\hf$ are scale-independent, their measured values can be used in the regime of our choice.

Eq.~\eqref{eq:Gamma} shows that we can measure $\Gamma$ in an agnostic way, without requiring any modeling of the redshift evolution of dark matter interactions: this is directly inferred from the data. From Eq.~\eqref{eq:Gamma}, we see that these measurements require: 1) measurements of $\hJ$ and $\hf$ at the same redshifts; 2) measurements of the redshift derivative of $\hf(z)$; and 3) measurements of the redshift evolution of the Hubble parameter $\HH(z)$. In our analysis we assume that 3) is known from background measurements of the expansion history of the Universe, namely from luminosity distance of supernovae, BAO and CMB measurements, and we fix it to a $\Lambda$CDM evolution. One can of course extend the method by reconstructing simultaneously $\HH(z)$ and $\Gamma(z)$.

\section*{Data availability statement}

The measurements of the growth rate of structure, $f\sigma_8$, from spectroscopic redshift surveys used in this analysis are publicly available and can be found in the respective papers. They are also listed in Table I of~\cite{Grimm:2024fui}. The measurements of the Weyl evolution $\hJ$ are available in~\cite{Tutusaus:2023aux}. Source data are provided in~\cite{codeFifthForce}.

\section*{Code availability statement}

The code used in this analysis is available at~\cite{codeFifthForce}. 
This analysis also made use of the public code \textsc{CAMB}~\cite{Lewis:1999bs}.

\begin{acknowledgments}
N.G.~and C.B.~acknowledge support from the European Research Council (ERC) under the European Union's Horizon 2020 research and innovation program (grant agreement No.~863929; project title ``Testing the law of gravity with novel large-scale structure observables''). I.T.~has been supported by the Ramon y Cajal fellowship (RYC2023-045531-I) funded by the State Research Agency of the Spanish Ministerio de Ciencia, Innovaci\'on y Universidades, MICIU/AEI/10.13039/501100011033/, and Social European Funds plus (FSE+). I.T.~also acknowledges support from the same ministry, via projects PID2019-11317GB, PID2022-141079NB, PID2022-138896NB; the European Research Executive Agency HORIZON-MSCA-2021-SE-01 Research and Innovation programme under the Marie Sk\l odowska-Curie grant agreement number 101086388 (LACEGAL) and the programme Unidad de Excelencia Mar\'{\i}a de Maeztu, project CEX2020-001058-M.
\end{acknowledgments}

\section*{Author Contributions Statement}

C.B. conceived the test and derived the equations. N.G. wrote the code to reconstruct the growth rate and constrain the fifth force. N.G. produced the Figures and Tables. N.G , C.B. and I.T. analysed the outputs, interpreted the results, and wrote the manuscript.

\section*{Competing Interests Statement}

The authors declare no competing interests.

\titleformat{\section}
  {\normalfont\fontsize{12}{14}\bfseries}{\thesection}{1em}{}
\titleformat{\subsection}
  {\normalfont\fontsize{10}{12}\bfseries}{\thesubsection}{1em}{}

\titlespacing*{\section}{0pt}{*2}{*1}
\titlespacing*{\subsection}{0pt}{*2}{*1}

\newpage

\section*{Supplementary information: Comparing the motion of dark matter and standard model particles on cosmological scales}
\appendix
\setcounter{section}{0}

\begin{supfigure}
    \centering
    \includegraphics[height = 6.7cm]{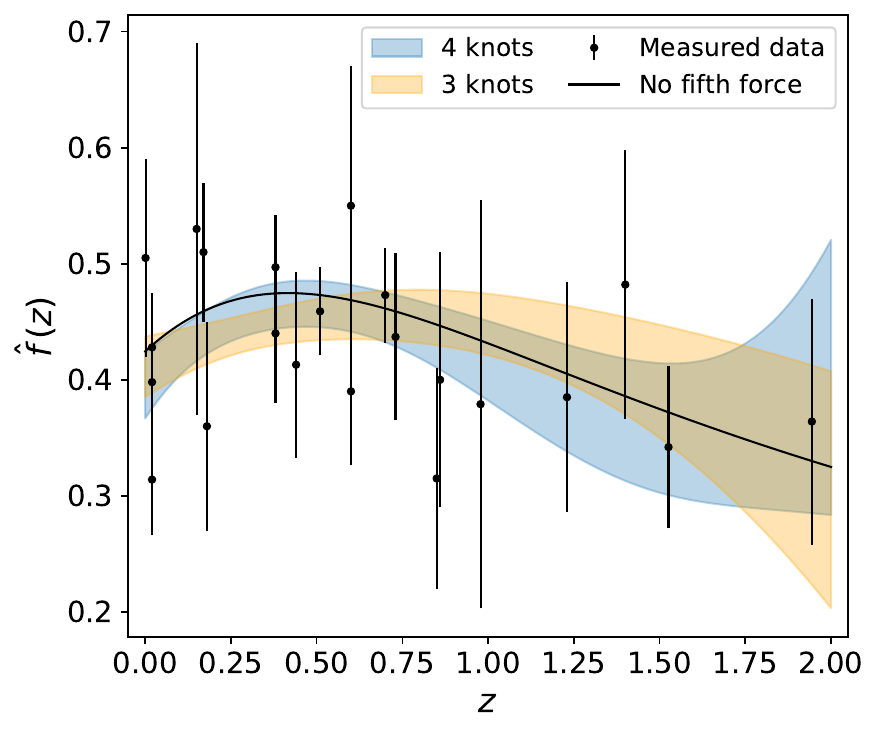}
    \includegraphics[height = 6.7cm]{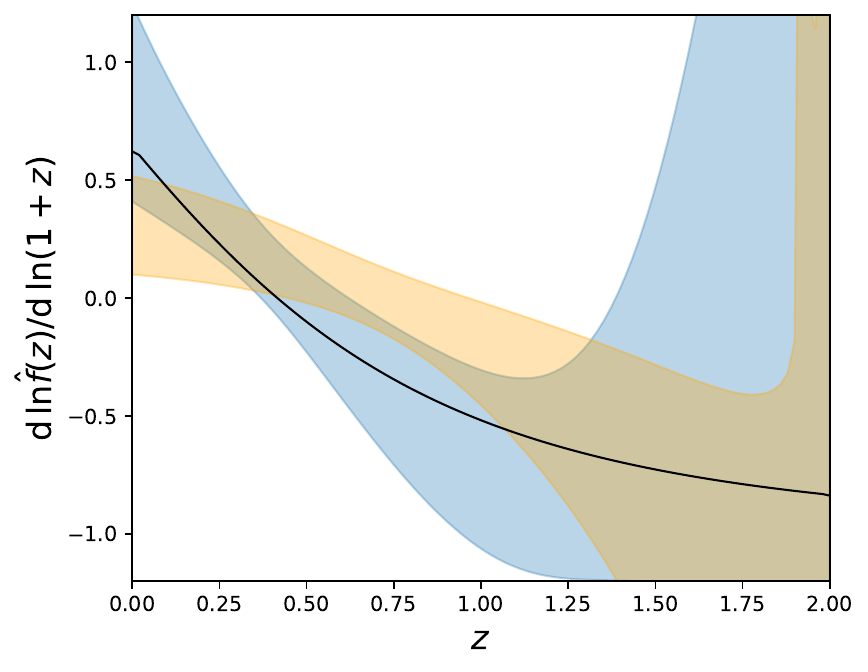}
    \caption{\justifying \textbf{Alternative reconstructions with current data} \textit{Left panel:} We show the spline reconstruction of $\hat{f}(z)$ with four knots in blue and with three knots in orange. In both cases the coloured region corresponds to $1\sigma$ uncertainties. \textit{Right panel:} We show the reconstruction of $\mathrm d \ln\hat{f}(z)/\mathrm d\ln(1+z)$, based on the spline interpolation of $\hat{f}$ for the two different cases. Source data are provided as a Source Data file~\cite{codeFifthForce}.} 
     \label{fig:f_curent_3_4}
\end{supfigure}

\begin{supfigure}
    \centering
    \includegraphics[width=.495\textwidth]{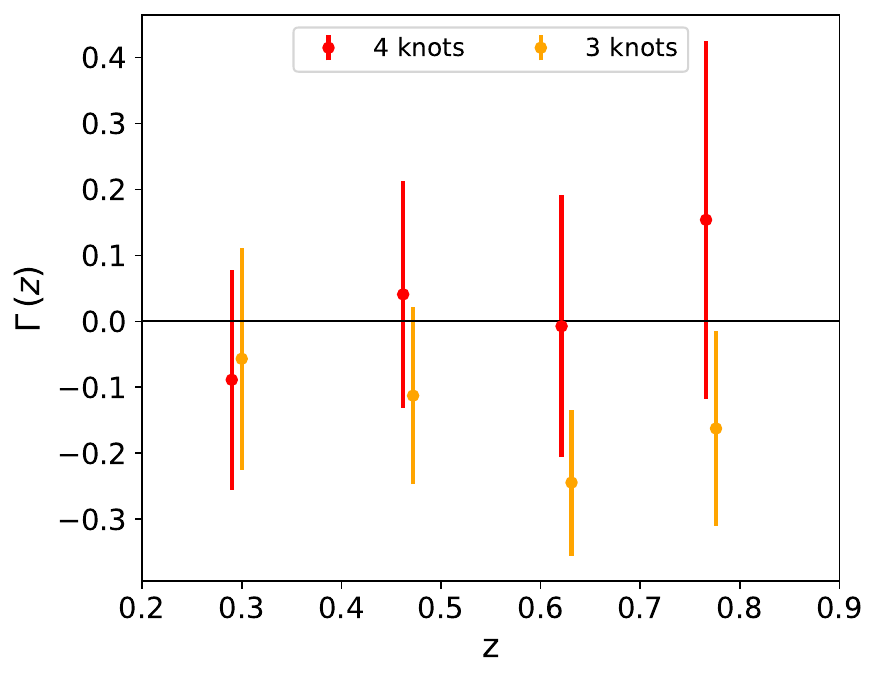}
    \caption{\justifying \textbf{Comparing constraints on $\Gamma$ for current data} We show the measured $\Gamma(z)$ with $1\sigma$ uncertainties at the four \texttt{MagLim} effective redshifts both for the case of four knots (red) and three knots (orange). Source data are provided as a Source Data file~\cite{codeFifthForce}.}\label{fig:Gamma_3_4}
\end{supfigure}

\section{Spline reconstruction with different numbers of redshift knots} \label{Appendix:Numbers_knots}

To reconstruct the function $\hat{f}(z)$, we have chosen a spline interpolation with four knots (current data) and five knots (future data) as our baseline case. This choice is based on the minimisation of the AIC, providing a balance between model fit (low chi-squared) and model complexity (number of parameters). Here, we further discuss how changing the number of knots affects the reconstruction.

\textit{Current surveys:} Supplementary Fig.~\ref{fig:f_curent_3_4} corresponds to Fig.~\ref{fig:f_current} in the main text, showing the reconstruction of $\hat{f}(z)$ (left panel) and its derivative (right panel) for current data. In addition to the baseline case with four knots in blue, we now also show in orange the case of three knots. This choice, corresponding effectively to a fit with a single second-degree polynomial, has a lower complexity at the expense of an increase of the chi-squared value and AIC. 
Supplementary Fig.~\ref{fig:Gamma_3_4} further shows the resulting values of $\Gamma(z)$ at the four \texttt{MagLim} effective redshifts in both cases. We see that, while both cases agree with each other at the $1\sigma$ level, the baseline case with four knots provides greater flexibility to capture the $\hf$ evolution. 
In particular, the case with three knots does not seem to have enough freedom to capture an evolution such as the $\Lambda$CDM one (in black), meaning that it could lead to spurious deviations in $\Gamma$ that arise due to underfitting rather than genuine physical effects. In addition to the two cases plotted here, we have also tested what happens when increasing to five or more knots: these more complex cases lead to wiggles in the reconstruction of $\hat{f}$ and particularly its derivative (due to the concatenation of several third-degree polynomials in the spline interpolation), and consequently to larger uncertainties in the results for $\Gamma(z)$.

\begin{supfigure}[t]
    \centering
    \includegraphics[height = 6.7cm]
    {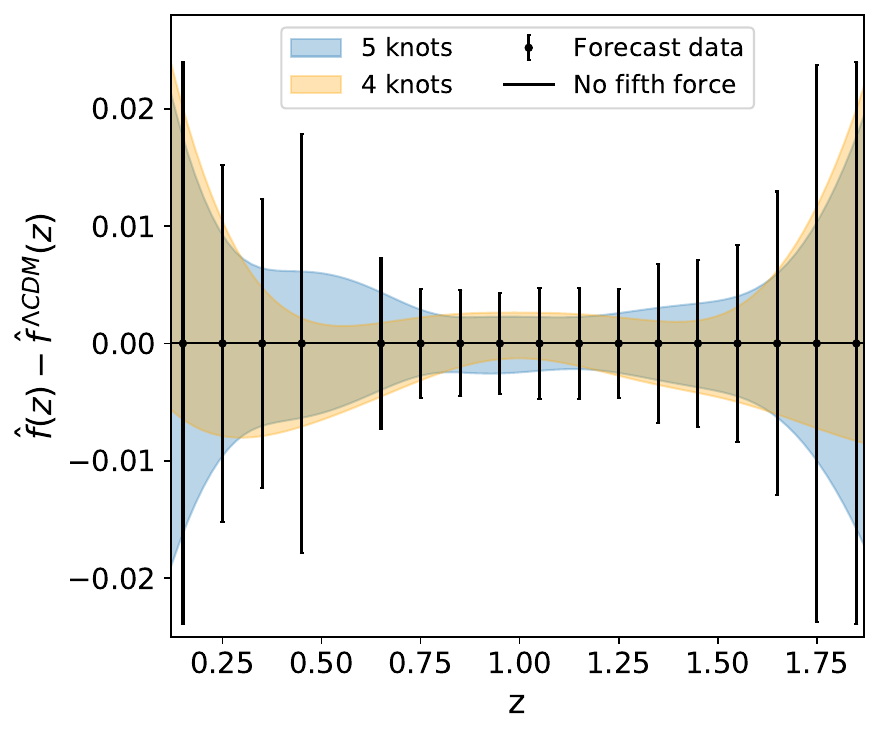}
    \includegraphics[height = 6.7cm]{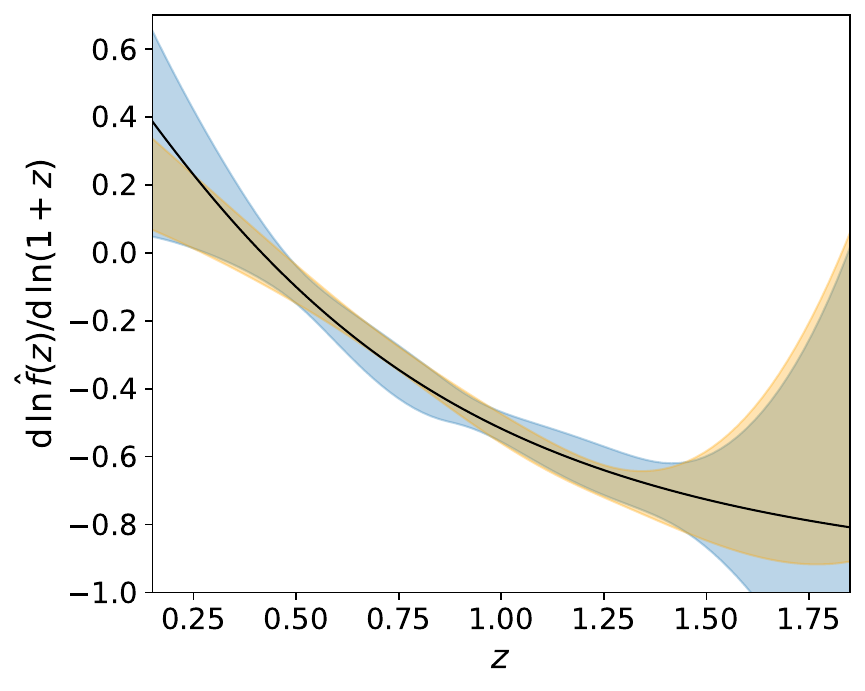}
    \caption{\justifying\textbf{Alternative reconstructions with future surveys} \textit{Left panel:} We show the spline reconstruction of $\hat{f}(z)$, for the case of future data, with five knots in blue and with four knots in orange. In both cases the coloured region corresponds to $1\sigma$ uncertainties. \textit{Right panel:} We show the reconstruction of $\mathrm d \ln\hat{f}(z)/\mathrm d\ln(1+z)$, based on the spline interpolation of $\hat{f}$ for the two different cases. Source data are provided as a Source Data file~\cite{codeFifthForce}.} 
     \label{fig:future_4_5}
\end{supfigure}

\begin{supfigure}[t]
    \centering
    \includegraphics[width=.495\textwidth]{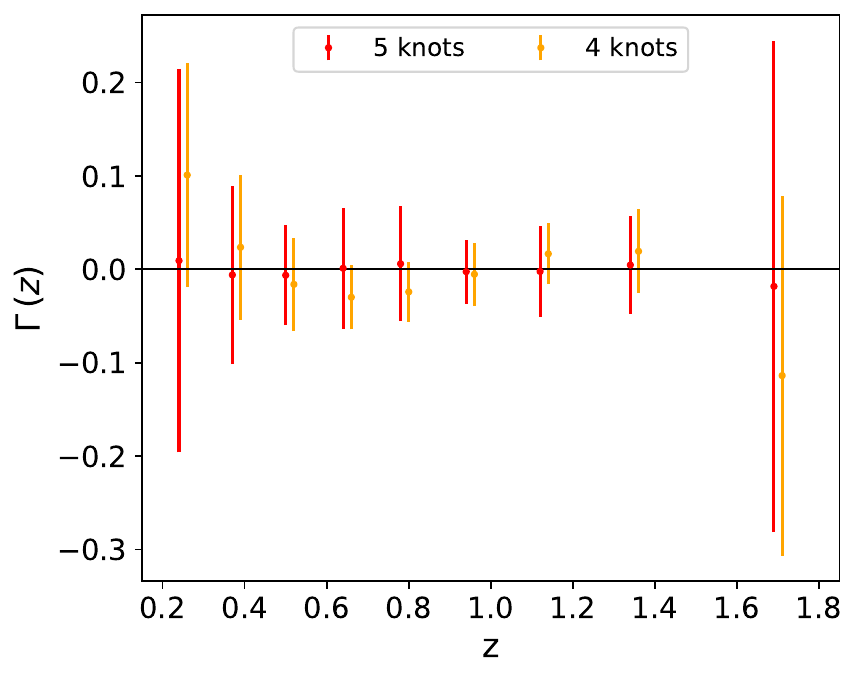}
    \caption{\justifying \textbf{Comparing constraints on $\Gamma$ for future surveys} We show the forecast $\Gamma(z)$ with $1\sigma$ uncertainties at nine LSST redshifts both for the case of five knots (red) and four knots (orange). Source data are provided as a Source Data file~\cite{codeFifthForce}.}\label{fig:Gamma_forecast_4_5}
\end{supfigure}

\textit{Forecasts with future surveys:} As discussed in the main text, our baseline choice for the reconstruction of $\hf$ from future data is five knots. Here we show the results for four knots, for which the chi-squared is larger while the AIC is reduced by a small amount (from 10.1 to 8.7) as the increase in the chi-squared value does not fully compensate the lower amount of free parameters. The comparison for the interpolation of $\hat{f}(z)$ is illustrated in the left panel of Supplementary Fig.~\ref{fig:future_4_5} (note that, to better visualise the difference between the two models, we subtract the fiducial value $\hat{f}^{\Lambda{\rm CDM}}(z)$). We see that the baseline case with five knots is better centred around the fiducial, and this behaviour is as well visible in the derivative of $\hat{f}(z)$ (right panel of Supplementary Fig.~\ref{fig:future_4_5}) and in the resulting values of $\Gamma(z)$ at the LSST redshifts (Supplementary Fig.~\ref{fig:Gamma_forecast_4_5}). Thus, as the five knots case shows a qualitatively better behaviour and the AIC favours four knots only mildly (a difference of $\Delta\rm{AIC}=1.4$ corresponds to a relative likelihood of 0.5 of the five knots case compared to the four knots case~\cite{Akaike:1974vps}), we choose to adopt five knots as our baseline case. We emphasise however that both cases are compatible with each other as well as the fiducial model on the $1\sigma$ level. Moreover, we note that, while for four knots the placement of the knots does not matter (since this case corresponds to a fit with a single third-degree polynomial assuming the standard \textit{not-a-knot} boundary condition), for five knots the choice of the middle knot needs to be specified. Applying spline interpolation with five knots, and enforcing the \textit{not-a-knot} boundary condition, results in two third-degree polynomials connected at the central knot, and the placement of this knot impacts the reconstruction.  We set the central knot to $z=0.86$, as the choice that minimises the chi-squared value and thus the AIC under the fixed assumption of five knots. The remaining knots then serve to fix the degrees of freedom of the polynomials, but their precise placements do not influence the reconstruction. Finally, we have also tested more complex cases including six knots or more (leading to a higher AIC). In these cases more wiggly features are introduced in the reconstruction of $\hat{f}$ and its derivative, and uncertainties on $\Gamma(z)$ increase.

\section{Covariance matrix for the parameter $\Gamma$ from current data}
\label{app:covariance}

Below, we list the covariance matrix for the four values of $\Gamma$ obtained from current data, as described in Section~\ref{sec:Results}:
\begin{equation}
    \mathrm{Cov}(\Gamma) = \begin{pmatrix}
0.0282 & 0.0173 & 0.0081 & 0.0026 \\
0.0173 & 0.0294 & 0.0282 & 0.0328 \\
0.0081 & 0.0282 & 0.0400 & 0.0478 \\
0.0026 & 0.0328 & 0.0478 & 0.0748
\end{pmatrix}
\,.
\end{equation}

\section{Velocity bias}
\label{app:velocity}

In our constraints we have assumed that the galaxy velocity equals the dark matter velocity: $V_{\rm g}=V_{\rm dm}$, i.e.\ that there is no velocity bias. Here we study how baryons can impact this assumption. We consider a fraction $x= \rho_{\rm dm}/(\rho_{\rm dm}+\rho_{\rm b})$ of dark matter subjected to a fifth force, and a fraction $1-x$ of baryons affected only by gravitational interaction. The baryon velocity, $V_{\rm b}$, and the dark matter velocity, $V_{\rm dm}$, obey the following equations
\begin{align}
&V'_{\rm b} + V_{\rm b}-\frac{k}{\HH}\Psi=0\, , \label{eq:Vb}\\   
&V'_{\rm dm} + (1+\theta)V_{\rm b}-\frac{k}{\HH}(1+\Gamma)\Psi=0\, .
\end{align}
The center of mass of the galaxy is given by a weighted average of the dark matter and baryon velocities
\begin{align}
V_{\rm g}=x V_{\rm dm}+(1-x)V_{\rm b}  \, .  
\end{align}
Assuming that $x$ is time-independent, we obtain the following modified Euler's equation for the galaxy velocity
\begin{align}
V'_{\rm g} + (1+\tilde{\theta})V_{\rm g}-\frac{k}{\HH}(1+\tilde{\Gamma})\Psi=0\, ,    
\end{align}
with $\tilde{\theta}=x (V_{\rm dm}/V_{\rm g})\,\theta$ and $\tilde{\Gamma}=x\Gamma $. For the models considered in this work, where $\theta$ is subdominant, we see that the only change to our constraints comes from ${\Gamma}$ being replaced by $x\Gamma$. The constraints are therefore degraded by a factor $x\simeq 0.8$.

The velocity field is measured from redshift-space distortions, i.e.\ from the Doppler shift experienced by light due to the motion of the emitter. In the case where baryons and dark matter do not move with the same velocity, due to the presence of a fifth force, one could wonder if it is truly $V_{\rm g}$ which is extracted from the multipoles of the correlation function, rather than $V_{\rm b}$. The light from galaxies is indeed emitted by baryons, not by dark matter. However, as shown in~\cite{Bonvin:2022tii} (see Section~\ref{sec:methods}), the key point is that the velocity of baryons with respect to the observer can be decomposed into the velocity of baryons with respect to the center of mass of the galaxies, plus the velocity of the center of mass with respect to the observer. When correlating distinct galaxies only the second part survives, showing that $V_{\rm g}$ is truly the quantity measured through redshift-space distortions.

\bibliography{fifthforce_DES}

@article{Lewis:1999bs,
    author = "Lewis, Antony and Challinor, Anthony and Lasenby, Anthony",
    title = "{Efficient computation of CMB anisotropies in closed FRW models}",
    eprint = "astro-ph/9911177",
    archivePrefix = "arXiv",
    doi = "10.1086/309179",
    journal = "Astrophys. J.",
    volume = "538",
    pages = "473--476",
    year = "2000"
}

@software{codeFifthForce,
  author = {Grimm, N.},
  title = {{Fifth force from fhat and Jhat}},
  url ={https://doi.org/10.5281/zenodo.17078450},
  version = {v0.1},
  DOI = {10.5281/zenodo.17078450},
  year = {2025}
}

@article{Bertschinger:2008zb,
    author = "Bertschinger, Edmund and Zukin, Phillip",
    title = "{Distinguishing Modified Gravity from Dark Energy}",
    eprint = "0801.2431",
    archivePrefix = "arXiv",
    primaryClass = "astro-ph",
    doi = "10.1103/PhysRevD.78.024015",
    journal = "Phys. Rev. D",
    volume = "78",
    pages = "024015",
    year = "2008"
}

@ARTICLE{2010PhRvD..81j4023P,
       author = {{Pogosian}, Levon and {Silvestri}, Alessandra and {Koyama}, Kazuya and {Zhao}, Gong-Bo},
        title = "{How to optimally parametrize deviations from general relativity in the evolution of cosmological perturbations}",
      journal = {\prd},
         year = 2010,
        month = may,
       volume = {81},
       number = {10},
          eid = {104023},
        pages = {104023},
          doi = {10.1103/PhysRevD.81.104023},
archivePrefix = {arXiv},
       eprint = {1002.2382},
 primaryClass = {astro-ph.CO},
       adsurl = {https://ui.adsabs.harvard.edu/abs/2010PhRvD..81j4023P}
}

@online{DESI2025,
  author       = {{Dark Energy Spectroscopic Instrument}},
  title        = {{DESI Collaboration}},
  year         = {},
  url          = {https://www.desi.lbl.gov},
}

@online{Euclid2025,
  author       = {{Euclid Survey}},
  title        = {{Euclid Consortium}},
  year         = {},
  url          = {https://www.euclid-ec.org},
}

@online{SKA2025,
  author       = {{Square Kilometer Array Observatory}},
  title        = {{SKAO}},
  year         = {},
  url          = {https://www.skao.int/en},
}

@online{LSST2025,
  author       = {{Rubin Observatory}},
  title        = {{LSST}},
  year         = {},
  url          = {https://rubinobservatory.org},
}

@online{SDSS2025,
  author       = {{Sloan Digital Sky Survey}},
  title        = {{SDSS Collaboration}},
  year         = {},
  url          = {https://www.sdss.org/},
}

@online{WiggleZ2025,
  author       = {{WiggleZ Dark Energy Survey}},
  title        = {{WiggleZ Collaboration}},
  year         = {},
  url          = {https://wigglez.swin.edu.au/site/forward.html},
}

@article{Kesden:2006zb,
    author = "Kesden, Michael and Kamionkowski, Marc",
    title = "{Galilean Equivalence for Galactic Dark Matter}",
    eprint = "astro-ph/0606566",
    archivePrefix = "arXiv",
    doi = "10.1103/PhysRevLett.97.131303",
    journal = "Phys. Rev. Lett.",
    volume = "97",
    pages = "131303",
    year = "2006"
}

@article{Akaike:1974vps,
    author = "Akaike, H.",
    title = "{A new look at the statistical model identification}",
    doi = "10.1109/TAC.1974.1100705",
    journal = "IEEE Trans. Automatic Control",
    volume = "19",
    number = "6",
    pages = "716--723",
    year = "1974"
}

@article{Castello:2022uuu,
    author = "Castello, Sveva and Grimm, Nastassia and Bonvin, Camille",
    title = "{Rescuing constraints on modified gravity using gravitational redshift in large-scale structure}",
    eprint = "2204.11507",
    archivePrefix = "arXiv",
    primaryClass = "astro-ph.CO",
    doi = "10.1103/PhysRevD.106.083511",
    journal = "Phys. Rev. D",
    volume = "106",
    number = "8",
    pages = "083511",
    year = "2022"
}

@article{DESI:2024hhd,
    author = "Adame, A. G. and others",
    collaboration = "DESI",
    title = "{DESI 2024 VII: cosmological constraints from the full-shape modeling of clustering measurements}",
    eprint = "2411.12022",
    archivePrefix = "arXiv",
    primaryClass = "astro-ph.CO",
    reportNumber = "FERMILAB-PUB-24-0854-PPD",
    doi = "10.1088/1475-7516/2025/07/028",
    journal = "JCAP",
    volume = "07",
    pages = "028",
    year = "2025"
}

@ARTICLE{2013ApJS..208...19H,
       author = {{Hinshaw}, G. and {Larson}, D. and {Komatsu}, E. and {Spergel}, D.~N. and {Bennett}, C.~L. and {Dunkley}, J. and {Nolta}, M.~R. and {Halpern}, M. and {Hill}, R.~S. and {Odegard}, N. and {Page}, L. and {Smith}, K.~M. and {Weiland}, J.~L. and {Gold}, B. and {Jarosik}, N. and {Kogut}, A. and {Limon}, M. and {Meyer}, S.~S. and {Tucker}, G.~S. and {Wollack}, E. and {Wright}, E.~L.},
        title = "{Nine-year Wilkinson Microwave Anisotropy Probe (WMAP) Observations: Cosmological Parameter Results}",
      journal = {Astrophys. J. Supp. Series},
         year = 2013,
        month = oct,
       volume = {208},
       number = {2},
          eid = {19},
        pages = {19},
          doi = {10.1088/0067-0049/208/2/19},
archivePrefix = {arXiv},
       eprint = {1212.5226},
 primaryClass = {astro-ph.CO},
       adsurl = {https://ui.adsabs.harvard.edu/abs/2013ApJS..208...19H}
}

@article{Peebles:1982ff,
    author = "Peebles, P. J. E.",
    editor = "Srednicki, M. A.",
    title = "{Large scale background temperature and mass fluctuations due to scale invariant primeval perturbations}",
    doi = "10.1086/183911",
    journal = "Astrophys. J. Lett.",
    volume = "263",
    pages = "L1--L5",
    year = "1982"
}

@article{Salucci:2018hqu,
    author = "Salucci, Paolo",
    title = "{The distribution of dark matter in galaxies}",
    eprint = "1811.08843",
    archivePrefix = "arXiv",
    primaryClass = "astro-ph.GA",
    doi = "10.1007/s00159-018-0113-1",
    journal = "Astron. Astrophys. Rev.",
    volume = "27",
    number = "1",
    pages = "2",
    year = "2019"
}

@article{Bertone:2016nfn,
    author = "Bertone, Gianfranco and Hooper, Dan",
    title = "{History of dark matter}",
    eprint = "1605.04909",
    archivePrefix = "arXiv",
    primaryClass = "astro-ph.CO",
    reportNumber = "FERMILAB-PUB-16-157-A",
    doi = "10.1103/RevModPhys.90.045002",
    journal = "Rev. Mod. Phys.",
    volume = "90",
    number = "4",
    pages = "045002",
    year = "2018"
}

@article{Davis:1985rj,
    author = "Davis, Marc and Efstathiou, George and Frenk, Carlos S. and White, Simon D. M.",
    editor = "Srednicki, M. A.",
    title = "{The Evolution of Large Scale Structure in a Universe Dominated by Cold Dark Matter}",
    reportNumber = "NSF-ITP-84-129",
    doi = "10.1086/163168",
    journal = "Astrophys. J.",
    volume = "292",
    pages = "371--394",
    year = "1985"
}

@article{Blumenthal:1984bp,
    author = "Blumenthal, George R. and Faber, S. M. and Primack, Joel R. and Rees, Martin J.",
    editor = "Srednicki, M. A.",
    title = "{Formation of Galaxies and Large Scale Structure with Cold Dark Matter}",
    reportNumber = "SLAC-PUB-3307",
    doi = "10.1038/311517a0",
    journal = "Nature",
    volume = "311",
    pages = "517--525",
    year = "1984"
}

@article{Chamings:2019kcl,
    author = "Chamings, Finlay Noble and Avgoustidis, Anastasios and Copeland, Edmund J. and Green, Anne M. and Pourtsidou, Alkistis",
    title = "{Understanding the suppression of structure formation from dark matter-dark energy momentum coupling}",
    eprint = "1912.09858",
    archivePrefix = "arXiv",
    primaryClass = "astro-ph.CO",
    doi = "10.1103/PhysRevD.101.043531",
    journal = "Phys. Rev. D",
    volume = "101",
    number = "4",
    pages = "043531",
    year = "2020"
}

@article{Amendola:2022vte,
    author = "Amendola, Luca and Pietroni, Massimo and Quartin, Miguel",
    title = "{Fisher matrix for the one-loop galaxy power spectrum: measuring expansion and growth rates without assuming a cosmological model}",
    eprint = "2205.00569",
    archivePrefix = "arXiv",
    primaryClass = "astro-ph.CO",
    doi = "10.1088/1475-7516/2022/11/023",
    journal = "JCAP",
    volume = "11",
    pages = "023",
    year = "2022"
}

@article{Schirra:2024rjq,
    author = "Schirra, Adrian P. and Quartin, Miguel and Amendola, Luca",
    title = "{A model-independent measurement of the expansion and growth rates from BOSS using the FreePower method}",
    eprint = "2406.15347",
    archivePrefix = "arXiv",
    primaryClass = "astro-ph.CO",
    doi = "10.1016/j.dark.2025.102033",
    journal = "Phys. Dark Univ.",
    volume = "49",
    pages = "102033",
    year = "2025"
}

@article{Castello:2024lhl,
    author = "Castello, Sveva and Zheng, Ziyang and Bonvin, Camille and Amendola, Luca",
    title = "{Testing the equivalence principle across the Universe: A model-independent approach with galaxy multitracing}",
    eprint = "2412.08627",
    archivePrefix = "arXiv",
    primaryClass = "astro-ph.CO",
    doi = "10.1103/1my7-zklj",
    journal = "Phys. Rev. D",
    volume = "111",
    number = "12",
    pages = "123559",
    year = "2025"
}

@article{Gaskins:2016cha,
    author = "Gaskins, Jennifer M.",
    title = "{A review of indirect searches for particle dark matter}",
    eprint = "1604.00014",
    archivePrefix = "arXiv",
    primaryClass = "astro-ph.HE",
    doi = "10.1080/00107514.2016.1175160",
    journal = "Contemp. Phys.",
    volume = "57",
    number = "4",
    pages = "496--525",
    year = "2016"
}

@article{Conrad:2017pms,
    author = "Conrad, Jan and Reimer, Olaf",
    title = "{Indirect dark matter searches in gamma and cosmic rays}",
    eprint = "1705.11165",
    archivePrefix = "arXiv",
    primaryClass = "astro-ph.HE",
    doi = "10.1038/nphys4049",
    journal = "Nature Phys.",
    volume = "13",
    number = "3",
    pages = "224--231",
    year = "2017"
}

@article{DESI:2024jis,
    author = "Adame, A. G. and others",
    collaboration = "DESI",
    title = "{DESI 2024 V: Full-Shape Galaxy Clustering from Galaxies and Quasars}",
    eprint = "2411.12021",
    archivePrefix = "arXiv",
    primaryClass = "astro-ph.CO",
    reportNumber = "FERMILAB-PUB-24-0847-PPD",
    month = "11",
    year = "2024"
}

@article{DESI:2016fyo,
    author = "Aghamousa, Amir and others",
    collaboration = "DESI",
    title = "{The DESI Experiment Part I: Science,Targeting, and Survey Design}",
    eprint = "1611.00036",
    archivePrefix = "arXiv",
    primaryClass = "astro-ph.IM",
    reportNumber = "FERMILAB-PUB-16-517-AE",
    month = "10",
    year = "2016"
}

@article{Huterer:2016uyq,
    author = "Huterer, Dragan and Shafer, Daniel and Scolnic, Daniel and Schmidt, Fabian",
    title = "{Testing $\Lambda$CDM at the lowest redshifts with SN Ia and galaxy velocities}",
    eprint = "1611.09862",
    archivePrefix = "arXiv",
    primaryClass = "astro-ph.CO",
    doi = "10.1088/1475-7516/2017/05/015",
    journal = "JCAP",
    volume = "05",
    pages = "015",
    year = "2017"
}

@article{Hudson:2012gt,
    author = "Hudson, Michael J. and Turnbull, Stephen J.",
    title = "{The growth rate of cosmic structure from peculiar velocities at low and high redshifts}",
    eprint = "1203.4814",
    archivePrefix = "arXiv",
    primaryClass = "astro-ph.CO",
    doi = "10.1088/2041-8205/751/2/L30",
    journal = "Astrophys. J. Lett.",
    volume = "751",
    pages = "L30",
    year = "2013"
}

@article{Turnbull:2011ty,
    author = "Turnbull, Stephen J. and Hudson, Michael J. and Feldman, Hume A. and Hicken, Malcolm and Kirshner, Robert P. and Watkins, Richard",
    title = "{Cosmic flows in the nearby universe from Type Ia Supernovae}",
    eprint = "1111.0631",
    archivePrefix = "arXiv",
    primaryClass = "astro-ph.CO",
    doi = "10.1111/j.1365-2966.2011.20050.x",
    journal = "Mon. Not. Roy. Astron. Soc.",
    volume = "420",
    pages = "447--454",
    year = "2012"
}

@article{Blake:2012pj,
    author = "Blake, Chris and others",
    title = "{The WiggleZ Dark Energy Survey: Joint measurements of the expansion and growth history at z \ensuremath{<} 1}",
    eprint = "1204.3674",
    archivePrefix = "arXiv",
    primaryClass = "astro-ph.CO",
    doi = "10.1111/j.1365-2966.2012.21473.x",
    journal = "Mon. Not. Roy. Astron. Soc.",
    volume = "425",
    pages = "405--414",
    year = "2012"
}

@article{eBOSS:2018yfg,
    author = "Zhao, Gong-Bo and others",
    collaboration = "eBOSS",
    title = "{The clustering of the SDSS-IV extended Baryon Oscillation Spectroscopic Survey DR14 quasar sample: a tomographic measurement of cosmic structure growth and expansion rate based on optimal redshift weights}",
    eprint = "1801.03043",
    archivePrefix = "arXiv",
    primaryClass = "astro-ph.CO",
    doi = "10.1093/mnras/sty2845",
    journal = "Mon. Not. Roy. Astron. Soc.",
    volume = "482",
    number = "3",
    pages = "3497--3513",
    year = "2019"
}

@article{Okumura:2015lvp,
    author = "Okumura, Teppei and others",
    title = "{The Subaru FMOS galaxy redshift survey (FastSound). IV. New constraint on gravity theory from redshift space distortions at $z\sim 1.4$}",
    eprint = "1511.08083",
    archivePrefix = "arXiv",
    primaryClass = "astro-ph.CO",
    doi = "10.1093/pasj/psw029",
    journal = "Publ. Astron. Soc. Jap.",
    volume = "68",
    number = "3",
    pages = "38",
    year = "2016"
}

@article{Song:2008qt,
    author = "Song, Yong-Seon and Percival, Will J.",
    title = "{Reconstructing the history of structure formation using Redshift Distortions}",
    eprint = "0807.0810",
    archivePrefix = "arXiv",
    primaryClass = "astro-ph",
    doi = "10.1088/1475-7516/2009/10/004",
    journal = "JCAP",
    volume = "10",
    pages = "004",
    year = "2009"
}

@article{Blake:2013nif,
    author = "Blake, Chris and others",
    title = "{Galaxy And Mass Assembly (GAMA): improved cosmic growth measurements using multiple tracers of large-scale structure}",
    eprint = "1309.5556",
    archivePrefix = "arXiv",
    primaryClass = "astro-ph.CO",
    doi = "10.1093/mnras/stt1791",
    journal = "Mon. Not. Roy. Astron. Soc.",
    volume = "436",
    pages = "3089",
    year = "2013"
}

@article{Davis:2010sw,
    author = "Davis, Marc and Nusser, Adi and Masters, Karen and Springob, Christopher and Huchra, John P. and Lemson, Gerard",
    title = "{Local Gravity versus Local Velocity: Solutions for $\beta$ and nonlinear bias}",
    eprint = "1011.3114",
    archivePrefix = "arXiv",
    primaryClass = "astro-ph.CO",
    doi = "10.1111/j.1365-2966.2011.18362.x",
    journal = "Mon. Not. Roy. Astron. Soc.",
    volume = "413",
    pages = "2906",
    year = "2011"
}

@article{Howlett:2017asq,
    author = {Howlett, Cullan and Staveley-Smith, Lister and Elahi, Pascal J. and Hong, Tao and Jarrett, Tom H. and Jones, D. Heath and Koribalski, B\"arbel S. and Macri, Lucas M. and Masters, Karen L. and Springob, Christopher M.},
    title = "{2MTF \textendash{} VI. Measuring the velocity power spectrum}",
    eprint = "1706.05130",
    archivePrefix = "arXiv",
    primaryClass = "astro-ph.CO",
    doi = "10.1093/mnras/stx1521",
    journal = "Mon. Not. Roy. Astron. Soc.",
    volume = "471",
    number = "3",
    pages = "3135--3151",
    year = "2017"
}

@article{ATLAS:2024fdw,
    author = "Aad, Georges and others",
    collaboration = "ATLAS",
    title = "{Exploration at the high-energy frontier: ATLAS Run~2 searches investigating the exotic jungle beyond the Standard Model}",
    eprint = "2403.09292",
    archivePrefix = "arXiv",
    primaryClass = "hep-ex",
    reportNumber = "CERN-EP-2024-075",
    doi = "10.1016/j.physrep.2024.10.001",
    journal = "Phys. Rept.",
    volume = "1116",
    pages = "301--385",
    year = "2025"
}

@article{ATLAS:2024itc,
    author = "Aad, Georges and others",
    collaboration = "ATLAS",
    title = "{ATLAS searches for additional scalars and exotic Higgs boson decays with the LHC Run~2 dataset}",
    eprint = "2405.04914",
    archivePrefix = "arXiv",
    primaryClass = "hep-ex",
    reportNumber = "CERN-EP-2024-094",
    doi = "10.1016/j.physrep.2024.09.002",
    journal = "Phys. Rept.",
    volume = "1116",
    pages = "184--260",
    year = "2025"
}

@article{ATLAS:2024lda,
    author = "Aad, Georges and others",
    collaboration = "ATLAS",
    title = "{The quest to discover supersymmetry at the ATLAS experiment}",
    eprint = "2403.02455",
    archivePrefix = "arXiv",
    primaryClass = "hep-ex",
    reportNumber = "CERN-EP-2024-056",
    doi = "10.1016/j.physrep.2024.09.010",
    journal = "Phys. Rept.",
    volume = "1116",
    pages = "261--300",
    year = "2025"
}

@article{FASER:2018eoc,
    author = "Ariga, Akitaka and others",
    collaboration = "FASER",
    title = "{FASER\textquoteright{}s physics reach for long-lived particles}",
    eprint = "1811.12522",
    archivePrefix = "arXiv",
    primaryClass = "hep-ph",
    reportNumber = "UCI-TR-2018-19, KYUSHU-RCAPP-2018-06",
    doi = "10.1103/PhysRevD.99.095011",
    journal = "Phys. Rev. D",
    volume = "99",
    number = "9",
    pages = "095011",
    year = "2019"
}

@article{Khoury:2003aq,
      author         = "Khoury, Justin and Weltman, Amanda",
      title          = "{Chameleon fields: Awaiting surprises for tests of
                        gravity in space}",
      journal        = "Phys. Rev. Lett.",
      volume         = "93",
      year           = "2004",
      pages          = "171104",
      doi            = "10.1103/PhysRevLett.93.171104",
      eprint         = "astro-ph/0309300",
      archivePrefix  = "arXiv",
      primaryClass   = "astro-ph",
      SLACcitation   = "%%CITATION = ASTRO-PH/0309300;%%"
}

@article{Hinterbichler:2010es,
      author         = "Hinterbichler, Kurt and Khoury, Justin",
      title          = "{Symmetron Fields: Screening Long-Range Forces Through
                        Local Symmetry Restoration}",
      journal        = "Phys. Rev. Lett.",
      volume         = "104",
      year           = "2010",
      pages          = "231301",
      doi            = "10.1103/PhysRevLett.104.231301",
      eprint         = "1001.4525",
      archivePrefix  = "arXiv",
      primaryClass   = "hep-th",
      SLACcitation   = "%%CITATION = ARXIV:1001.4525;%%"
}

@ARTICLE{RobertsonBAHAMAS,
   author = {{Robertson}, A. and David Harvey and {Massey}, R. and {Eke}, V. and 
	{McCarthy}, I.~G. and {Jauzac}, M. and {Li}, B. and {Schaye}, J.
	},
    title = "{Observable tests of self-interacting dark matter in galaxy clusters: cosmological simulations with SIDM and baryons}",
  journal = {\mnras},
archivePrefix = "arXiv",
   eprint = {1810.05649},
     year = 2019,
    month = sep,
   volume = 488,
    pages = {3646-3662},
      doi = {10.1093/mnras/stz1815},
   adsurl = {https://ui.adsabs.harvard.edu/abs/2019MNRAS.488.3646R}
}

@article{Eckert:2022qia,
    author = "Eckert, D. and Ettori, S. and Robertson, A. and Massey, R. and Pointecouteau, E. and Harvey, D. and McCarthy, I. G.",
    title = "{Constraints on dark matter self-interaction from the internal density profiles of X-COP galaxy clusters}",
    eprint = "2205.01123",
    archivePrefix = "arXiv",
    primaryClass = "astro-ph.CO",
    doi = "10.1051/0004-6361/202243205",
    journal = "Astron. Astrophys.",
    volume = "666",
    pages = "A41",
    year = "2022"
}

@ARTICLE{harveyIA,
doi = {10.1093/mnras/stab1741},  
     journal = "\mnras",
eprint = {2104.02093},
	year = 2021,
	month = {jun},
	publisher = {Oxford University Press ({OUP})},
	volume = {506},
	number = {1},
	pages = {441--451},
	author = {David Harvey and Nora Elisa Chisari and Andrew Robertson and Ian G McCarthy},
	title = {The impact of self-interacting dark matter on the intrinsic alignments of galaxies},
}

@article{Rodriguez:2021urv,
    author = {Rodr\'\i{}guez, Alexandre Brea and Chobanova, Veronika and Cid Vidal, Xabier and Soli\~no, Sa\'ul L\'opez and Santos, Diego Mart\'\i{}nez and Momb\"acher, Titus and Prouv\'e, Claire and Fern\'andez, Emilio Xos\'e Rodr\'\i{}guez and V\'azquez Sierra, Carlos},
    title = "{Prospects on searches for baryonic Dark Matter produced in b-hadron decays at LHCb}",
    eprint = "2106.12870",
    archivePrefix = "arXiv",
    primaryClass = "hep-ph",
    doi = "10.1140/epjc/s10052-021-09762-w",
    journal = "Eur. Phys. J. C",
    volume = "81",
    number = "11",
    pages = "964",
    year = "2021"
}

@article{PhysRevLett.124.041801,
    author = "Aaij, Roel and others",
    collaboration = "LHCb",
    title = "{Search for $A'\to\mu^+\mu^-$ Decays}",
    eprint = "1910.06926",
    archivePrefix = "arXiv",
    primaryClass = "hep-ex",
    reportNumber = "LHCb-PAPER-2019-031, CERN-EP-2019-212",
    doi = "10.1103/PhysRevLett.124.041801",
    journal = "Phys. Rev. Lett.",
    volume = "124",
    number = "4",
    pages = "041801",
    year = "2020"
}

@article{CRESST:2019jnq,
    author = "Abdelhameed, A. H. and others",
    collaboration = "CRESST",
    title = "{First results from the CRESST-III low-mass dark matter program}",
    eprint = "1904.00498",
    archivePrefix = "arXiv",
    primaryClass = "astro-ph.CO",
    doi = "10.1103/PhysRevD.100.102002",
    journal = "Phys. Rev. D",
    volume = "100",
    number = "10",
    pages = "102002",
    year = "2019"
}

@article{Behnke:2016lsk,
    author = "Behnke, E. and others",
    title = "{Final Results of the PICASSO Dark Matter Search Experiment}",
    eprint = "1611.01499",
    archivePrefix = "arXiv",
    primaryClass = "hep-ex",
    doi = "10.1016/j.astropartphys.2017.02.005",
    journal = "Astropart. Phys.",
    volume = "90",
    pages = "85--92",
    year = "2017"
}

@article{DarkSide-50:2022qzh,
    author = "Agnes, P. and others",
    collaboration = "DarkSide-50",
    title = "{Search for low-mass dark matter WIMPs with 12~ton-day exposure of DarkSide-50}",
    eprint = "2207.11966",
    archivePrefix = "arXiv",
    primaryClass = "hep-ex",
    reportNumber = "FERMILAB-PUB-22-589-ND-PPD-SCD",
    doi = "10.1103/PhysRevD.107.063001",
    journal = "Phys. Rev. D",
    volume = "107",
    number = "6",
    pages = "063001",
    year = "2023"
}

@article{LZ:2022lsv,
    author = "Aalbers, J. and others",
    collaboration = "LZ",
    title = "{First Dark Matter Search Results from the LUX-ZEPLIN (LZ) Experiment}",
    eprint = "2207.03764",
    archivePrefix = "arXiv",
    primaryClass = "hep-ex",
    doi = "10.1103/PhysRevLett.131.041002",
    journal = "Phys. Rev. Lett.",
    volume = "131",
    number = "4",
    pages = "041002",
    year = "2023"
}

@article{CMS:2024zqs,
    author = "Hayrapetyan, Aram and others",
    collaboration = "CMS",
    title = "{Dark sector searches with the CMS experiment}",
    eprint = "2405.13778",
    archivePrefix = "arXiv",
    primaryClass = "hep-ex",
    reportNumber = "CMS-EXO-23-005, CERN-EP-2024-106",
    doi = "10.1016/j.physrep.2024.09.013",
    journal = "Phys. Rept.",
    volume = "1115",
    pages = "448--569",
    year = "2025"
}

@article{Creminelli:2013nua,
    author = "Creminelli, Paolo and Gleyzes, J\'er\^ome and Hui, Lam and Simonovi\'c, Marko and Vernizzi, Filippo",
    title = "{Single-Field Consistency Relations of Large Scale Structure. Part III: Test of the Equivalence Principle}",
    eprint = "1312.6074",
    archivePrefix = "arXiv",
    primaryClass = "astro-ph.CO",
    doi = "10.1088/1475-7516/2014/06/009",
    journal = "JCAP",
    volume = "06",
    pages = "009",
    year = "2014"
}

@article{Kehagias:2013rpa,
    author = "Kehagias, Alexandros and Nore\~na, Jorge and Perrier, Hideki and Riotto, Antonio",
    title = "{Consequences of Symmetries and Consistency Relations in the Large-Scale Structure of the Universe for Non-local bias and Modified Gravity}",
    eprint = "1311.0786",
    archivePrefix = "arXiv",
    primaryClass = "astro-ph.CO",
    doi = "10.1016/j.nuclphysb.2014.03.020",
    journal = "Nucl. Phys. B",
    volume = "883",
    pages = "83--106",
    year = "2014"
}

@article{Wang:2023tjj,
    author = "Wang, Zhuangfei and Mirpoorian, Seyed Hamidreza and Pogosian, Levon and Silvestri, Alessandra and Zhao, Gong-Bo",
    title = "{New MGCAMB tests of gravity with CosmoMC and Cobaya}",
    eprint = "2305.05667",
    archivePrefix = "arXiv",
    primaryClass = "astro-ph.CO",
    doi = "10.1088/1475-7516/2023/08/038",
    journal = "JCAP",
    volume = "08",
    pages = "038",
    year = "2023"
}

@article{Grimm:2024fui,
    author = "Grimm, Nastassia and Bonvin, Camille and Tutusaus, Isaac",
    title = "{Testing General Relativity through the EG Statistic Using the Weyl Potential and Galaxy Velocities}",
    eprint = "2403.13709",
    archivePrefix = "arXiv",
    primaryClass = "astro-ph.CO",
    doi = "10.1103/PhysRevLett.133.211004",
    journal = "Phys. Rev. Lett.",
    volume = "133",
    number = "21",
    pages = "211004",
    year = "2024"
}

@article{Castello:2024jmq,
    author = "Castello, Sveva and Wang, Zhuangfei and Dam, Lawrence and Bonvin, Camille and Pogosian, Levon",
    title = "{Disentangling modified gravity from a dark force with gravitational redshift}",
    eprint = "2404.09379",
    archivePrefix = "arXiv",
    primaryClass = "astro-ph.CO",
    doi = "10.1103/PhysRevD.110.103523",
    journal = "Phys. Rev. D",
    volume = "110",
    number = "10",
    pages = "103523",
    year = "2024"
}

@article{Tutusaus:2023aux,
    author = "Tutusaus, Isaac and Bonvin, Camille and Grimm, Nastassia",
    title = "{Measurement of the Weyl potential evolution from the first three years of dark energy survey data}",
    eprint = "2312.06434",
    archivePrefix = "arXiv",
    primaryClass = "astro-ph.CO",
    doi = "10.1038/s41467-024-53363-6",
    journal = "Nature Commun.",
    volume = "15",
    number = "1",
    pages = "9295",
    year = "2024"
}

@article{Bottaro:2023wkd,
    author = "Bottaro, Salvatore and Castorina, Emanuele and Costa, Marco and Redigolo, Diego and Salvioni, Ennio",
    title = "{Unveiling dark forces with the Large Scale Structure of the Universe}",
    eprint = "2309.11496",
    archivePrefix = "arXiv",
    primaryClass = "astro-ph.CO",
    doi = "10.1103/PhysRevLett.132.201002",
    journal = "Phys. Rev. Lett.",
    volume = "132",
    number = "20",
    pages = "201002",
    year = "2024"
}

@article{Bonvin:2018ckp,
    author = "Bonvin, Camille and Fleury, Pierre",
    title = "{Testing the equivalence principle on cosmological scales}",
    eprint = "1803.02771",
    archivePrefix = "arXiv",
    primaryClass = "astro-ph.CO",
    doi = "10.1088/1475-7516/2018/05/061",
    journal = " \jcap",
    volume = "05",
    pages = "061",
    year = "2018"
}

@article{Sobral-Blanco:2022oel,
    author = "Sobral-Blanco, Daniel and Bonvin, Camille",
    title = "{Measuring the distortion of time with relativistic effects in large-scale structure}",
    eprint = "2205.02567",
    archivePrefix = "arXiv",
    primaryClass = "astro-ph.CO",
    doi = "10.1093/mnrasl/slac124",
    journal = " \mnras",
    volume = "519",
    number = "1",
    pages = "L39--L44",
    year = "2022"
}

@article{Bonvin:2022tii,
    author = "Bonvin, Camille and Pogosian, Levon",
    title = "{Modified Einstein versus Modified Euler for Dark Matter}",
    eprint = "2209.03614",
    archivePrefix = "arXiv",
    primaryClass = "astro-ph.CO",
    doi = "10.1038/s41550-023-02003-y",
    journal = "Nature Astron.",
    volume = "7",
    number = "9",
    pages = "1127--1134",
    year = "2023"
}

@article{Tutusaus:2022cab,
doi = {10.1103/physrevd.107.083526},  
	year = 2023,
	month = {apr},
	publisher = {American Physical Society ({APS})},
	volume = {107},
	number = {8},
    author = "Tutusaus, Isaac and Sobral-Blanco, Daniel and Bonvin, Camille",
    title = "{Combining gravitational lensing and gravitational redshift to measure the anisotropic stress with future galaxy surveys}",
    eprint = "2209.08987",
    archivePrefix = "arXiv",
    primaryClass = "astro-ph.CO",
    journal = {\prd },
}

@article{Desmond:2020gzn,
    author = "Desmond, Harry and Ferreira, Pedro G.",
    title = "{Galaxy morphology rules out astrophysically relevant Hu-Sawicki $f(R)$ gravity}",
    eprint = "2009.08743",
    archivePrefix = "arXiv",
    primaryClass = "astro-ph.CO",
    doi = "10.1103/PhysRevD.102.104060",
    journal = " \prd",
    volume = "102",
    number = "10",
    pages = "104060",
    year = "2020"
}

@article{Planck:2018vyg,
    author = {N. Aghanim and Y. Akrami and M. Ashdown and J. Aumont and C. Baccigalupi and M. Ballardini and A. J. Banday and R. B. Barreiro and N. Bartolo and S. Basak and R. Battye and K. Benabed and J.-P. Bernard and M. Bersanelli and P. Bielewicz and J. J. Bock and J. R. Bond and J. Borrill and F. R. Bouchet and F. Boulanger and M. Bucher and C. Burigana and R. C. Butler and E. Calabrese and J.-F. Cardoso and J. Carron and A. Challinor and H. C. Chiang and J. Chluba and L. P. L. Colombo and C. Combet and D. Contreras and B. P. Crill and F. Cuttaia and P. de Bernardis and G. de Zotti and J. Delabrouille and J.-M. Delouis and E. Di Valentino and J. M. Diego and O. Dor{\'{e}} and M. Douspis and A. Ducout and X. Dupac and S. Dusini and G. Efstathiou and F. Elsner and T. A. En{\ss}lin and H. K. Eriksen and Y. Fantaye and M. Farhang and J. Fergusson and R. Fernandez-Cobos and F. Finelli and F. Forastieri and M. Frailis and A. A. Fraisse and E. Franceschi and A. Frolov and S. Galeotta and S. Galli and K. Ganga and R. T. G{\'{e}}nova-Santos and M. Gerbino and T. Ghosh and J. Gonz{\'{a}}lez-Nuevo and K. M. G{\'{o}}rski and S. Gratton and A. Gruppuso and J. E. Gudmundsson and J. Hamann and W. Handley and F. K. Hansen and D. Herranz and S. R. Hildebrandt and E. Hivon and Z. Huang and A. H. Jaffe and W. C. Jones and A. Karakci and E. Keihänen and R. Keskitalo and K. Kiiveri and J. Kim and T. S. Kisner and L. Knox and N. Krachmalnicoff and M. Kunz and H. Kurki-Suonio and G. Lagache and J.-M. Lamarre and A. Lasenby and M. Lattanzi and C. R. Lawrence and M. Le Jeune and P. Lemos and J. Lesgourgues and F. Levrier and A. Lewis and M. Liguori and P. B. Lilje and M. Lilley and V. Lindholm and M. L{\'{o}}pez-Caniego and P. M. Lubin and Y.-Z. Ma and J. F. Mac{\'{\i}}as-P{\'{e}}rez and G. Maggio and D. Maino and N. Mandolesi and A. Mangilli and A. Marcos-Caballero and M. Maris and P. G. Martin and M. Martinelli and E. Mart{\'{\i}}nez-Gonz{\'{a}}lez and S. Matarrese and N. Mauri and J. D. McEwen and P. R. Meinhold and A. Melchiorri and A. Mennella and M. Migliaccio and M. Millea and S. Mitra and M.-A. Miville-Desch{\^{e}}nes and D. Molinari and L. Montier and G. Morgante and A. Moss and P. Natoli and H. U. N{\o}rgaard-Nielsen and L. Pagano and D. Paoletti and B. Partridge and G. Patanchon and H. V. Peiris and F. Perrotta and V. Pettorino and F. Piacentini and L. Polastri and G. Polenta and J.-L. Puget and J. P. Rachen and M. Reinecke and M. Remazeilles and A. Renzi and G. Rocha and C. Rosset and G. Roudier and J. A. Rubi{\~{n}}o-Mart{\'{\i}}n and B. Ruiz-Granados and L. Salvati and M. Sandri and M. Savelainen and D. Scott and E. P. S. Shellard and C. Sirignano and G. Sirri and L. D. Spencer and R. Sunyaev and A.-S. Suur-Uski and J. A. Tauber and D. Tavagnacco and M. Tenti and L. Toffolatti and M. Tomasi and T. Trombetti and L. Valenziano and J. Valiviita and B. Van Tent and L. Vibert and P. Vielva and F. Villa and N. Vittorio and B. D. Wandelt and I. K. Wehus and M. White and S. D. M. White and A. Zacchei and A. Zonca},
    collaboration = "Planck",
    title = "{Planck 2018 results. VI. Cosmological parameters}",
    eprint = "1807.06209",
    archivePrefix = "arXiv",
    primaryClass = "astro-ph.CO",
    doi = "10.1051/0004-6361/201833910",
    journal = " \aap",
    volume = "641",
    pages = "A6",
    year = "2020",
    note = "[Erratum: \aap 652, C4 (2021)]"
}

@article{Pourtsidou:2013nha,
    author = "Pourtsidou, A. and Skordis, C. and Copeland, E. J.",
    title = "{Models of dark matter coupled to dark energy}",
    eprint = "1307.0458",
    archivePrefix = "arXiv",
    primaryClass = "astro-ph.CO",
    doi = "10.1103/PhysRevD.88.083505",
    journal = " \prd",
    volume = "88",
    number = "8",
    pages = "083505",
    year = "2013"
}

@article{Euclid:2019clj,
    author = {A. Blanchard and S. Camera and C. Carbone and V. F. Cardone and S. Casas and S. Clesse and S. Ili{\'{c}
} and M. Kilbinger and T. Kitching and M. Kunz and F. Lacasa and E. Linder and E. Majerotto and K. Markovi{\v{c}} and M. Martinelli and V. Pettorino and A. Pourtsidou and Z. Sakr and A. G. S{\'{a}}nchez and D. Sapone and I. Tutusaus and S. Yahia-Cherif and V. Yankelevich and S. Andreon and H. Aussel and A. Balaguera-Antol{\'{\i}}nez and M. Baldi and S. Bardelli and R. Bender and A. Biviano and D. Bonino and A. Boucaud and E. Bozzo and E. Branchini and S. Brau-Nogue and M. Brescia and J. Brinchmann and C. Burigana and R. Cabanac and V. Capobianco and A. Cappi and J. Carretero and C. S. Carvalho and R. Casas and F. J. Castander and M. Castellano and S. Cavuoti and A. Cimatti and R. Cledassou and C. Colodro-Conde and G. Congedo and C. J. Conselice and L. Conversi and Y. Copin and L. Corcione and J. Coupon and H. M. Courtois and M. Cropper and A. Da Silva and S. de la Torre and D. Di Ferdinando and F. Dubath and F. Ducret and C. A. J. Duncan and X. Dupac and S. Dusini and G. Fabbian and M. Fabricius and S. Farrens and P. Fosalba and S. Fotopoulou and N. Fourmanoit and M. Frailis and E. Franceschi and P. Franzetti and M. Fumana and S. Galeotta and W. Gillard and B. Gillis and C. Giocoli and P. G{\'{o}}mez-Alvarez and J. Graci{\'{a}}-Carpio and F. Grupp and L. Guzzo and H. Hoekstra and F. Hormuth and H. Israel and K. Jahnke and E. Keihanen and S. Kermiche and C. C. Kirkpatrick and R. Kohley and B. Kubik and H. Kurki-Suonio and S. Ligori and P. B. Lilje and I. Lloro and D. Maino and E. Maiorano and O. Marggraf and N. Martinet and F. Marulli and R. Massey and E. Medinaceli and S. Mei and Y. Mellier and B. Metcalf and J. J. Metge and G. Meylan and M. Moresco and L. Moscardini and E. Munari and R. C. Nichol and S. Niemi and A. A. Nucita and C. Padilla and S. Paltani and F. Pasian and W. J. Percival and S. Pires and G. Polenta and M. Poncet and L. Pozzetti and G. D. Racca and F. Raison and A. Renzi and J. Rhodes and E. Romelli and M. Roncarelli and E. Rossetti and R. Saglia and P. Schneider and V. Scottez and A. Secroun and G. Sirri and L. Stanco and J.-L. Starck and F. Sureau and P. Tallada-Cresp{\'{\i}} and D. Tavagnacco and A. N. Taylor and M. Tenti and I. Tereno and R. Toledo-Moreo and F. Torradeflot and L. Valenziano and T. Vassallo and G. A. Verdoes Kleijn and M. Viel and Y. Wang and A. Zacchei and J. Zoubian and E. Zucca},
    collaboration = "Euclid",
    title = "{Euclid preparation: VII. Forecast validation for Euclid cosmological probes}",
    eprint = "1910.09273",
    archivePrefix = "arXiv",
    primaryClass = "astro-ph.CO",
    doi = "10.1051/0004-6361/202038071",
    journal = " \aap",
    volume = "642",
    pages = "A191",
    year = "2020"
}

@article{BOSS:2016ntk,
    author = {Siddharth Satpathy and Shadab Alam and Shirley Ho and Martin White and Neta A. Bahcall and Florian Beutler and Joel R. Brownstein and Chia-Hsun Chuang and Daniel J. Eisenstein and Jan Niklas Grieb and Francisco Kitaura and Matthew D. Olmstead and Will J. Percival and Salvador Salazar-Albornoz and Ariel G. S{\'{a}}nchez and Hee-Jong Seo and Daniel Thomas and Jeremy L. Tinker and Rita Tojeiro},
    collaboration = "BOSS",
    title = "{The clustering of galaxies in the completed SDSS-III Baryon Oscillation Spectroscopic Survey: On the measurement of growth rate using galaxy correlation functions}",
    eprint = "1607.03148",
    archivePrefix = "arXiv",
    primaryClass = "astro-ph.CO",
    doi = "10.1093/mnras/stx883",
    journal = " \mnras",
    volume = "469",
    number = "2",
    pages = "1369--1382",
    year = "2017"
}

@article{DES:2018ufa,
    author = {T. M. C. Abbott and F. B. Abdalla and S. Avila and M. Banerji and E. Baxter and K. Bechtol and M. R. Becker and E. Bertin and J. Blazek and S. L. Bridle and D. Brooks and D. Brout and D. L. Burke and A. Campos and A. Carnero Rosell and M. Carrasco Kind and J. Carretero and F. J. Castander and R. Cawthon and C. Chang and A. Chen and M. Crocce and C. E. Cunha and L. N. da Costa and C. Davis and J. De Vicente and J. DeRose and S. Desai and E. Di Valentino and H. T. Diehl and J. P. Dietrich and S. Dodelson and P. Doel and A. Drlica-Wagner and T. F. Eifler and J. Elvin-Poole and A. E. Evrard and E. Fernandez and A. Fert{\'{e}} and B. Flaugher and P. Fosalba and J. Frieman and J. Garc{\'{\i}}a-Bellido and E. Gaztanaga and D. W. Gerdes and T. Giannantonio and D. Gruen and R. A. Gruendl and J. Gschwend and G. Gutierrez and W. G. Hartley and D. L. Hollowood and K. Honscheid and B. Hoyle and D. Huterer and B. Jain and T. Jeltema and M. W. G. Johnson and M. D. Johnson and A. G. Kim and E. Krause and K. Kuehn and N. Kuropatkin and O. Lahav and S. Lee and P. Lemos and C. D. Leonard and T. S. Li and A. R. Liddle and M. Lima and H. Lin and M. A. G. Maia and J. L. Marshall and P. Martini and F. Menanteau and C. J. Miller and R. Miquel and V. Miranda and J. J. Mohr and J. Muir and R. C. Nichol and B. Nord and R. L. C. Ogando and A. A. Plazas and M. Raveri and R. P. Rollins and A. K. Romer and A. Roodman and R. Rosenfeld and S. Samuroff and E. Sanchez and V. Scarpine and R. Schindler and M. Schubnell and D. Scolnic and L. F. Secco and S. Serrano and I. Sevilla-Noarbe and M. Smith and M. Soares-Santos and F. Sobreira and E. Suchyta and M. E. C. Swanson and G. Tarle and D. Thomas and M. A. Troxel and V. Vikram and A. R. Walker and N. Weaverdyck and R. H. Wechsler and J. Weller and B. Yanny and Y. Zhang and J. Zuntz},
    collaboration = "DES",
    title = "{Dark Energy Survey Year 1 Results: Constraints on Extended Cosmological Models from Galaxy Clustering and Weak Lensing}",
    eprint = "1810.02499",
    archivePrefix = "arXiv",
    primaryClass = "astro-ph.CO",
    reportNumber = "FERMILAB-PUB-18-507-PPD",
    doi = "10.1103/PhysRevD.99.123505",
    journal = " \prd",
    volume = "99",
    number = "12",
    pages = "123505",
    year = "2019"
}

@article{Tulin:2017ara,
    author = "Tulin, Sean and Yu, Hai-Bo",
    title = "{Dark Matter Self-interactions and Small Scale Structure}",
    eprint = "1705.02358",
    archivePrefix = "arXiv",
    primaryClass = "hep-ph",
    doi = "10.1016/j.physrep.2017.11.004",
    journal = "Phys. Rept.",
    volume = "730",
    pages = "1--57",
    year = "2018"
}

@article{Spergel:1999mh,
    author = "Spergel, David N. and Steinhardt, Paul J.",
    title = "{Observational evidence for selfinteracting cold dark matter}",
    eprint = "astro-ph/9909386",
    archivePrefix = "arXiv",
    doi = "10.1103/PhysRevLett.84.3760",
    journal = " \prl",
    volume = "84",
    pages = "3760--3763",
    year = "2000"
}

@article{Diacoumis:2018ezi,
    author = "Diacoumis, James A. D. and Wong, Yvonne Y. Y.",
    title = "{On the prior dependence of cosmological constraints on some dark matter interactions}",
    eprint = "1811.11408",
    archivePrefix = "arXiv",
    primaryClass = "astro-ph.CO",
    doi = "10.1088/1475-7516/2019/05/025",
    journal = " \jcap",
    volume = "05",
    pages = "025",
    year = "2019"
}

@article{Schewtschenko:2015rno,
    author = "Schewtschenko, J. A. and Baugh, C. M. and Wilkinson, R. J. and B\oe{}hm, C. and Pascoli, S. and Sawala, T.",
    title = "{Dark matter\textendash{}radiation interactions: the structure of Milky Way satellite galaxies}",
    eprint = "1512.06774",
    archivePrefix = "arXiv",
    primaryClass = "astro-ph.CO",
    reportNumber = "DURAST-2016-0001, IPPP-16-03",
    doi = "10.1093/mnras/stw1078",
    journal = " \mnras",
    volume = "461",
    number = "3",
    pages = "2282--2287",
    year = "2016"
}

@article{Barkana:2018lgd,
    author = "Barkana, Rennan",
    title = "{Possible interaction between baryons and dark-matter particles revealed by the first stars}",
    eprint = "1803.06698",
    archivePrefix = "arXiv",
    primaryClass = "astro-ph.CO",
    doi = "10.1038/nature25791",
    journal = "Nature",
    volume = "555",
    number = "7694",
    pages = "71--74",
    year = "2018"
}

@article{Diamanti:2012tg,
    author = "Diamanti, Roberta and Giusarma, Elena and Mena, Olga and Archidiacono, Maria and Melchiorri, Alessandro",
    title = "{Dark Radiation and interacting scenarios}",
    eprint = "1212.6007",
    archivePrefix = "arXiv",
    primaryClass = "astro-ph.CO",
    doi = "10.1103/PhysRevD.87.063509",
    journal = " \prd",
    volume = "87",
    number = "6",
    pages = "063509",
    year = "2013"
}

@article{Amendola:2007rr,
      author         = "Amendola, Luca and Kunz, Martin and Sapone, Domenico",
      title          = "{Measuring the dark side (with weak lensing)}",
      journal        = "JCAP",
      volume         = "0804",
      year           = "2008",
      pages          = "013",
      doi            = "10.1088/1475-7516/2008/04/013",
      eprint         = "0704.2421",
      archivePrefix  = "arXiv",
      primaryClass   = "astro-ph",
      SLACcitation   = "%%CITATION = ARXIV:0704.2421;%%"
}

@article{Daniel:2008et,
      author         = "Daniel, Scott F. and Caldwell, Robert R. and Cooray,
                        Asantha and Melchiorri, Alessandro",
      title          = "{Large Scale Structure as a Probe of Gravitational Slip}",
      journal        = "Phys. Rev.",
      volume         = "D77",
      year           = "2008",
      pages          = "103513",
      doi            = "10.1103/PhysRevD.77.103513",
      eprint         = "0802.1068",
      archivePrefix  = "arXiv",
      primaryClass   = "astro-ph",
      SLACcitation   = "%%CITATION = ARXIV:0802.1068;%%"
}

@Inbook{Hamilton:1997zq,
author="Hamilton, A. J. S.",
title="Linear Redshift Distortions: A Review",
bookTitle="The Evolving Universe: Selected Topics on Large-Scale Structure and on the Properties of Galaxies",
eprint = "astro-ph/9708102",
archivePrefix = "arXiv",
year="1998",
publisher="Springer Netherlands",
address="Dordrecht",
pages="185--275",
isbn="978-94-011-4960-0",
doi="10.1007/978-94-011-4960-0_17"
}

@article{Kaiser:1987qv,
    author = "Kaiser, N.",
    title = "{Clustering in real space and in redshift space}",
    journal = "\mnras",
    volume = "227",
    pages = "1--27",
    year = "1987",
    doi = {10.1093/mnras/227.1.1}
}

@article{KiDS:2020suj,
    author = {Marika Asgari and Chieh-An Lin and Benjamin Joachimi and Benjamin Giblin and Catherine Heymans and Hendrik Hildebrandt and Arun Kannawadi and Benjamin Stölzner and Tilman Tröster and Jan Luca van den Busch and Angus H. Wright and Maciej Bilicki and Chris Blake and Jelte de Jong and Andrej Dvornik and Thomas Erben and Fedor Getman and Henk Hoekstra and Fabian Köhlinger and Konrad Kuijken and Lance Miller and Mario Radovich and Peter Schneider and HuanYuan Shan and Edwin Valentijn},
    collaboration = "KiDS collaboration",
    title = "{KiDS-1000 Cosmology: Cosmic shear constraints and comparison between two point statistics}",
    eprint = "2007.15633",
    archivePrefix = "arXiv",
    primaryClass = "astro-ph.CO",
    doi = "10.1051/0004-6361/202039070",
    journal = "\aap",
    volume = "645",
    pages = "A104",
    year = "2021"
}

@article{Pezzotta:2016gbo,
    	author = {A. Pezzotta and S. de la Torre and J. Bel and B. R. Granett and L. Guzzo and J. A. Peacock and B. Garilli and M. Scodeggio and M. Bolzonella and U. Abbas and C. Adami and D. Bottini and A. Cappi and O. Cucciati and I. Davidzon and P. Franzetti and A. Fritz and A. Iovino and J. Krywult and V. Le Brun and O. Le F{\`{e}
}vre and D. Maccagni and K. Ma{\l}ek and F. Marulli and M. Polletta and A. Pollo and L. A. M. Tasca and R. Tojeiro and D. Vergani and A. Zanichelli and S. Arnouts and E. Branchini and J. Coupon and G. De Lucia and J. Koda and O. Ilbert and F. Mohammad and T. Moutard and L. Moscardini},
    title = "{The VIMOS Public Extragalactic Redshift Survey (VIPERS): The growth of structure at $0.5 < z < 1.2$ from redshift-space distortions in the clustering of the PDR-2 final sample}",
    eprint = "1612.05645",
    archivePrefix = "arXiv",
    primaryClass = "astro-ph.CO",
    doi = "10.1051/0004-6361/201630295",
    journal = "\aap",
    volume = "604",
    pages = "A33",
    year = "2017"
}

@article{eBOSS:2020yzd,
    author = {Shadab Alam and Marie Aubert and Santiago Avila and Christophe Balland and Julian E. Bautista and Matthew A. Bershady and Dmitry Bizyaev and Michael R. Blanton and Adam S. Bolton and Jo Bovy and Jonathan Brinkmann and Joel R. Brownstein and Etienne Burtin and Sol{\`{e}}ne Chabanier and Michael J. Chapman and Peter Doohyun Choi and Chia-Hsun Chuang and Johan Comparat and Marie-Claude Cousinou and Andrei Cuceu and Kyle S. Dawson and Sylvain de la Torre and Arnaud de Mattia and Victoria de Sainte Agathe and H{\'{e}}lion du Mas des Bourboux and Stephanie Escoffier and Thomas Etourneau and James Farr and Andreu Font-Ribera and Peter M. Frinchaboy and Sebastien Fromenteau and H{\'{e}}ctor Gil-Mar{\'{\i}}n and Jean-Marc Le Goff and Alma X. Gonzalez-Morales and Violeta Gonzalez-Perez and Kathleen Grabowski and Julien Guy and Adam J. Hawken and Jiamin Hou and Hui Kong and James Parker and Mark Klaene and Jean-Paul Kneib and Sicheng Lin and Daniel Long and Brad W. Lyke and Axel de la Macorra and Paul Martini and Karen Masters and Faizan G. Mohammad and Jeongin Moon and Eva-Maria Mueller and Andrea Mu{\~{n}}oz-Guti{\'{e}}rrez and Adam D. Myers and Seshadri Nadathur and Richard Neveux and Jeffrey A. Newman and Pasquier Noterdaeme and Audrey Oravetz and Daniel Oravetz and Nathalie Palanque-Delabrouille and Kaike Pan and Romain Paviot and Will J. Percival and Ignasi P{\'{e}}rez-R{\`{a}}fols and Patrick Petitjean and Matthew M. Pieri and Abhishek Prakash and Anand Raichoor and Corentin Ravoux and Mehdi Rezaie and James Rich and Ashley J. Ross and Graziano Rossi and Rossana Ruggeri and Vanina Ruhlmann-Kleider and Ariel G. S{\'{a}}nchez and F. Javier S{\'{a}}nchez and Jos{\'{e}} R. S{\'{a}}nchez-Gallego and Conor Sayres and Donald P. Schneider and Hee-Jong Seo and Arman Shafieloo and An{\v{z}}e Slosar and Alex Smith and Julianna Stermer and Amelie Tamone and Jeremy L. Tinker and Rita Tojeiro and Mariana Vargas-Maga{\~{n}}a and Andrei Variu and Yuting Wang and Benjamin A. Weaver and Anne-Marie Weijmans and Christophe Y{\`{e}}che and Pauline Zarrouk and Cheng Zhao and Gong-Bo Zhao and Zheng Zheng},
    collaboration = "eBOSS collaboration",
    title = "{Completed {SDSS-IV} extended Baryon Oscillation Spectroscopic Survey: Cosmological implications from two decades of spectroscopic surveys at the Apache Point Observatory}",
    eprint = "2007.08991",
    archivePrefix = "arXiv",
    primaryClass = "astro-ph.CO",
    doi = "10.1103/PhysRevD.103.083533",
    journal = "\prd",
    volume = "103",
    number = "8",
    pages = "083533",
    year = "2021"
}

@article{Gleyzes:2015pma,
    author = "Gleyzes, J\'er\^ome and Langlois, David and Mancarella, Michele and Vernizzi, Filippo",
    title = "{Effective Theory of Interacting Dark Energy}",
    eprint = "1504.05481",
    archivePrefix = "arXiv",
    primaryClass = "astro-ph.CO",
    doi = "10.1088/1475-7516/2015/08/054",
    journal = "\jcap",
    volume = "08",
    pages = "054",
    year = "2015"
}

@article{Costa:2014pba,
    author = "Costa, Andr\'e A. and Olivari, Lucas C. and Abdalla, E.",
    title = "{Quintessence with Yukawa Interaction}",
    eprint = "1411.3660",
    archivePrefix = "arXiv",
    primaryClass = "astro-ph.CO",
    doi = "10.1103/PhysRevD.92.103501",
    journal = "\prd",
    volume = "92",
    number = "10",
    pages = "103501",
    year = "2015"
}

@article{Pettorino_2012,
doi = {10.1103/physrevd.86.103507},
year = 2012,
month = {nov},
publisher = {American Physical Society ({APS})},
volume = {86},
number = {10},
author = {Valeria Pettorino and Luca Amendola and Carlo Baccigalupi and Claudia Quercellini},
title = {Constraints on coupled dark energy using {CMB} data from {WMAP} and South Pole Telescope},
eprint = {1207.3293},
journal = {\prd}
}

@article{Archidiacono:2019wdp,
    author = "Archidiacono, Maria and Hooper, Deanna C. and Murgia, Riccardo and Bohr, Sebastian and Lesgourgues, Julien and Viel, Matteo",
    title = "{Constraining Dark Matter-Dark Radiation interactions with CMB, BAO, and Lyman-$\alpha$}",
    eprint = "1907.01496",
    archivePrefix = "arXiv",
    primaryClass = "astro-ph.CO",
    doi = "10.1088/1475-7516/2019/10/055",
    journal = "\jcap",
    volume = "10",
    pages = "055",
    year = "2019"
}

@article{Buen-Abad:2015ova,
    author = "Buen-Abad, Manuel A. and Marques-Tavares, Gustavo and Schmaltz, Martin",
    title = "{Non-Abelian dark matter and dark radiation}",
    eprint = "1505.03542",
    archivePrefix = "arXiv",
    primaryClass = "hep-ph",
    doi = "10.1103/PhysRevD.92.023531",
    journal = "\prd",
    volume = "92",
    number = "2",
    pages = "023531",
    year = "2015"
}

@article{Archidiacono:2022iuu,
    author = "Archidiacono, Maria and Castorina, Emanuele and Redigolo, Diego and Salvioni, Ennio",
    title = "{Unveiling dark fifth forces with linear cosmology}",
    eprint = "2204.08484",
    archivePrefix = "arXiv",
    primaryClass = "astro-ph.CO",
    reportNumber = "CERN-TH-2022-066",
    doi = "10.1088/1475-7516/2022/10/074",
    journal = "\jcap",
    volume = "10",
    pages = "074",
    year = "2022"
}

@article{DES:2021wwk,
    author = {{Abbott}, T.~M.~C. and {Aguena}, M. and {Alarcon}, A. and {Allam}, S. and {Alves}, O. and {Amon}, A. and {Andrade-Oliveira}, F. and {Annis}, J. and {Avila}, S. and {Bacon}, D. and {Baxter}, E. and {Bechtol}, K. and {Becker}, M.~R. and {Bernstein}, G.~M. and {Bhargava}, S. and {Birrer}, S. and {Blazek}, J. and {Brandao-Souza}, A. and {Bridle}, S.~L. and {Brooks}, D. and {Buckley-Geer}, E. and {Burke}, D.~L. and {Camacho}, H. and {Campos}, A. and {Carnero Rosell}, A. and {Carrasco Kind}, M. and {Carretero}, J. and {Castander}, F.~J. and {Cawthon}, R. and {Chang}, C. and {Chen}, A. and {Chen}, R. and {Choi}, A. and {Conselice}, C. and {Cordero}, J. and {Costanzi}, M. and {Crocce}, M. and {da Costa}, L.~N. and {da Silva Pereira}, M.~E. and {Davis}, C. and {Davis}, T.~M. and {De Vicente}, J. and {DeRose}, J. and {Desai}, S. and {Di Valentino}, E. and {Diehl}, H.~T. and {Dietrich}, J.~P. and {Dodelson}, S. and {Doel}, P. and {Doux}, C. and {Drlica-Wagner}, A. and {Eckert}, K. and {Eifler}, T.~F. and {Elsner}, F. and {Elvin-Poole}, J. and {Everett}, S. and {Evrard}, A.~E. and {Fang}, X. and {Farahi}, A. and {Fernandez}, E. and {Ferrero}, I. and {Fert{\'e}}, A. and {Fosalba}, P. and {Friedrich}, O. and {Frieman}, J. and {Garc{\'\i}a-Bellido}, J. and {Gatti}, M. and {Gaztanaga}, E. and {Gerdes}, D.~W. and {Giannantonio}, T. and {Giannini}, G. and {Gruen}, D. and {Gruendl}, R.~A. and {Gschwend}, J. and {Gutierrez}, G. and {Harrison}, I. and {Hartley}, W.~G. and {Herner}, K. and {Hinton}, S.~R. and {Hollowood}, D.~L. and {Honscheid}, K. and {Hoyle}, B. and {Huff}, E.~M. and {Huterer}, D. and {Jain}, B. and {James}, D.~J. and {Jarvis}, M. and {Jeffrey}, N. and {Jeltema}, T. and {Kovacs}, A. and {Krause}, E. and {Kron}, R. and {Kuehn}, K. and {Kuropatkin}, N. and {Lahav}, O. and {Leget}, P. -F. and {Lemos}, P. and {Liddle}, A.~R. and {Lidman}, C. and {Lima}, M. and {Lin}, H. and {MacCrann}, N. and {Maia}, M.~A.~G. and {Marshall}, J.~L. and {Martini}, P. and {McCullough}, J. and {Melchior}, P. and {Mena-Fern{\'a}ndez}, J. and {Menanteau}, F. and {Miquel}, R. and {Mohr}, J.~J. and {Morgan}, R. and {Muir}, J. and {Myles}, J. and {Nadathur}, S. and {Navarro-Alsina}, A. and {Nichol}, R.~C. and {Ogando}, R.~L.~C. and {Omori}, Y. and {Palmese}, A. and {Pandey}, S. and {Park}, Y. and {Paz-Chinch{\'o}n}, F. and {Petravick}, D. and {Pieres}, A. and {Plazas Malag{\'o}n}, A.~A. and {Porredon}, A. and {Prat}, J. and {Raveri}, M. and {Rodriguez-Monroy}, M. and {Rollins}, R.~P. and {Romer}, A.~K. and {Roodman}, A. and {Rosenfeld}, R. and {Ross}, A.~J. and {Rykoff}, E.~S. and {Samuroff}, S. and {S{\'a}nchez}, C. and {Sanchez}, E. and {Sanchez}, J. and {Sanchez Cid}, D. and {Scarpine}, V. and {Schubnell}, M. and {Scolnic}, D. and {Secco}, L.~F. and {Serrano}, S. and {Sevilla-Noarbe}, I. and {Sheldon}, E. and {Shin}, T. and {Smith}, M. and {Soares-Santos}, M. and {Suchyta}, E. and {Swanson}, M.~E.~C. and {Tabbutt}, M. and {Tarle}, G. and {Thomas}, D. and {To}, C. and {Troja}, A. and {Troxel}, M.~A. and {Tucker}, D.~L. and {Tutusaus}, I. and {Varga}, T.~N. and {Walker}, A.~R. and {Weaverdyck}, N. and {Wechsler}, R. and {Weller}, J. and {Yanny}, B. and {Yin}, B. and {Zhang}, Y. and {Zuntz}, J.},
    collaboration = "DES collaboration",
    title = "{Dark Energy Survey Year 3 results: Cosmological constraints from galaxy clustering and weak lensing}",
    eprint = "2105.13549",
    archivePrefix = "arXiv",
    primaryClass = "astro-ph.CO",
    reportNumber = "FERMILAB-PUB-21-221-AE, DES-2020-0617",
    doi = "10.1103/PhysRevD.105.023520",
    journal = "\prd",
    volume = "105",
    number = "2",
    pages = "023520",
    year = "2022"
}

@article{DES:2022ygi,
    author={T. M. C. Abbott and M. Aguena and A. Alarcon and O. Alves and A. Amon and J. Annis and S. Avila and D. Bacon and E. Baxter and K. Bechtol and M. R. Becker and G. M. Bernstein and S. Birrer and J. Blazek and S. Bocquet and A. Brandao-Souza and S. L. Bridle and D. Brooks and D. L. Burke and H. Camacho and A. Campos and A. Carnero Rosell and M. Carrasco Kind and J. Carretero and F. J. Castander and R. Cawthon and C. Chang and A. Chen and R. Chen and A. Choi and C. Conselice and J. Cordero and M. Costanzi and M. Crocce and L. N. da Costa and M. E. S. Pereira and C. Davis and T. M. Davis and J. DeRose and S. Desai and E. Di Valentino and H. T. Diehl and S. Dodelson and P. Doel and C. Doux and A. Drlica-Wagner and K. Eckert and T. F. Eifler and F. Elsner and J. Elvin-Poole and S. Everett and X. Fang and A. Farahi and I. Ferrero and A. Ferté and B. Flaugher and P. Fosalba and D. Friedel and O. Friedrich and J. Frieman and J. García-Bellido and M. Gatti and L. Giani and T. Giannantonio and G. Giannini and D. Gruen and R. A. Gruendl and J. Gschwend and G. Gutierrez and N. Hamaus and I. Harrison and W. G. Hartley and K. Herner and S. R. Hinton and D. L. Hollowood and K. Honscheid and H. Huang and E. M. Huff and D. Huterer and B. Jain and D. J. James and M. Jarvis and N. Jeffrey and T. Jeltema and A. Kovacs and E. Krause and K. Kuehn and N. Kuropatkin and O. Lahav and S. Lee and P. -F. Leget and P. Lemos and C. D. Leonard and A. R. Liddle and M. Lima and H. Lin and N. MacCrann and J. L. Marshall and J. McCullough and J. Mena-Fernández and F. Menanteau and R. Miquel and V. Miranda and J. J. Mohr and J. Muir and J. Myles and S. Nadathur and A. Navarro-Alsina and R. C. Nichol and R. L. C. Ogando and Y. Omori and A. Palmese and S. Pandey and Y. Park and M. Paterno and F. Paz-Chinchón and W. J. Percival and A. Pieres and A. A. Plazas Malagón and A. Porredon and J. Prat and M. Raveri and M. Rodriguez-Monroy and P. Rogozenski and R. P. Rollins and A. K. Romer and A. Roodman and R. Rosenfeld and A. J. Ross and E. S. Rykoff and S. Samuroff and C. Sánchez and E. Sanchez and J. Sanchez and D. Sanchez Cid and V. Scarpine and D. Scolnic and L. F. Secco and I. Sevilla-Noarbe and E. Sheldon and T. Shin and M. Smith and M. Soares-Santos and E. Suchyta and M. Tabbutt and G. Tarle and D. Thomas and C. To and A. Troja and M. A. Troxel and I. Tutusaus and T. N. Varga and M. Vincenzi and A. R. Walker and N. Weaverdyck and R. H. Wechsler and J. Weller and B. Yanny and B. Yin and Y. Zhang and J. Zuntz},
    collaboration = "DES collaboration",
    title = "{Dark Energy Survey Year 3 Results: Constraints on extensions to $\Lambda${CDM} with weak lensing and galaxy clustering}",
    eprint = "2207.05766",
    archivePrefix = "arXiv",
    primaryClass = "astro-ph.CO",
    journal = " \prd",
    volume = "107",
    number = "8",
    pages = "083504",
    doi = "10.1103/PhysRevD.107.083504",
    year = "2023",
}

@article{Perenon:2021uom,
    author = "Perenon, Louis and Martinelli, Matteo and Ili\'c, St\'ephane and Maartens, Roy and Lochner, Michelle and Clarkson, Chris",
    title = "{Multi-tasking the growth of cosmological structures}",
    eprint = "2105.01613",
    archivePrefix = "arXiv",
    primaryClass = "astro-ph.CO",
    doi = "10.1016/j.dark.2021.100898",
    journal = "Phys. Dark Univ.",
    volume = "34",
    pages = "100898",
    year = "2021"
}
\bibliographystyle{naturemag}

\end{document}